\documentclass[twocolumn,aps,prd,floatfix,preprintnumbers,a4paper,nofootinbib,superscriptaddress,10pt]{revtex4-2}

\usepackage{amssymb,amsmath,verbatim,mathtools,needspace,enumitem,etoolbox,graphicx,microtype}
\usepackage{multirow}
\usepackage[dvipsnames, usenames]{xcolor}
\usepackage{bigints}
\usepackage[top=0.85in,
  bottom=1.in,
  left=0.75in,
  right=0.75in]{geometry}
\usepackage{subcaption}
\definecolor{linkcolor}{rgb}{0.0,0.4,0.4}
\definecolor{citecolor}{rgb}{.7,.3,.5}
\usepackage[colorlinks=true, linkcolor=linkcolor, citecolor=citecolor, pdfusetitle]{hyperref}

\usepackage[all]{hypcap}
\usepackage[T1]{fontenc}
\usepackage[utf8]{inputenc}
\usepackage{graphicx,amssymb,amsmath,amsthm,amsfonts,epsfig,times,natbib}
\usepackage[normalem]{ulem}
\usepackage{multirow}
\usepackage{import}
\usepackage{array}

\usepackage[titletoc,title]{appendix}

\usepackage{graphicx}				
\usepackage{placeins}
\usepackage{subcaption}
\usepackage{ragged2e} 
\DeclareCaptionFormat{myformat}{\justifying#1#2#3}
\usepackage{verbatim}
\usepackage{rotating}
\usepackage{comment}
\usepackage{txfonts}

\usepackage{epstopdf}
\usepackage{tensor}
\usepackage{amsbsy}
\usepackage{bm,url}
\usepackage{float}
\usepackage{dcolumn}
\usepackage{latexsym}
\usepackage{rotating}
\usepackage{enumerate}
\newcommand{\msun}{\ensuremath{M_\odot}}

\def\be{\begin{equation}}
\def\ee{\end{equation}}
\def\bea{\begin{eqnarray}}
\def\eea{\end{eqnarray}}

\newcommand{\ba}{\begin{align}}
\newcommand{\ea}{\end{align}}

\newcommand{\deco}{\texttt{PhenomDECO}}
\newcommand{\bbh}{ \texttt{PhenomD}}
\newcommand{\xphm}{ \texttt{PhenomXPHM}}

\newcommand{\AEI}{Max Planck Institute for Gravitational Physics (Albert Einstein Institute),
Callinstrasse 38, D-30167 Hannover, Germany}
\newcommand{\Leibniz}{Leibniz University Hannover, 30167 Hannover, Germany}
\newcommand{\Portsmouth}{Institute of Cosmology and Gravitation, University of Portsmouth, Portsmouth, PO1 3FX, UK}
\newcommand{\Cardiff}{Gravity Exploration Institute, Cardiff University, Cardiff, United Kingdom}

\begin{document}
\title{Compactness Inference in Gravitational-Wave Mergers with PhenomDECO: Catalog Benchmarks and Robustness Diagnostics}
\author{Shrobana Ghosh}
\email{shrobana.ghosh@aei.mpg.de}
\affiliation{\AEI}
\affiliation{\Leibniz}
\author{Charlie Hoy}
\email{charlie.hoy@port.ac.uk}
\affiliation{\Portsmouth}
\author{Mark Hannam} 
\affiliation{\Cardiff}
\author{Frank Ohme}
\affiliation{\AEI}
\affiliation{\Leibniz}

\begin{abstract}
Several gravitational wave (GW) observations have been identified as binary black hole (BBH) mergers, including systems with component masses that challenge typical formation scenarios. These observations motivate broader tests of whether the detected sources are consistent with this interpretation. We address this question using~\deco~, an existing phenomenological extension of a BBH waveform model that uses an effective compactness parameter to characterize departures from the expected merger morphology. Applying this model to all high-significance BBH events from GWTC-3, we establish~\deco~as a robust test of the nature of compact binaries. In preliminary analyses we identify three recurring effective compactness-posterior morphologies: (i) near-Gaussian peaks consistent with the BBH expectation $C\sim0.5$, observed for 60\% of events; (ii) posteriors with additional high-compactness support $(C\ge0.8)$; and (iii) dominant low-compactness modes near $C\sim0.15$ for $\sim 20\%$ of cases. For the latter, we find that the low-compactness modes disappear when the data in --  particularly in the Livingston detector --  is analyzed from a higher starting frequency, indicating sensitivity to low-frequency noise artefacts.We further use time--frequency residuals, computed after subtracting maximum-likelihood BBH and~\deco~waveforms from the strain data, as a complementary check on whether apparent residual structure is better described by the compactness deformation. Having established a robust analysis procedure, we conclude that all of the GWTC-3 observations that we have considered are indeed consistent with BBH sources. The exception is the high-mass GW231123 signal, where we need to analyse data from \emph{both} detectors above 50Hz to remove a low-compactness mode. This study shows that low-frequency data treatment is crucial before attributing apparent deviations from BBH expectations to exotic physics, and it provides a benchmark for interpreting compactness-based tests of merger morphology in current and future GW detections.

\end{abstract}

\maketitle

\section{Introduction}

The rapidly growing catalog of compact binary coalescences (CBCs)~\cite{GWTC3,Nitz_2023,PhysRevD.101.083030,Olsen:2022pin,Mehta:2023zlk} has widened the possibility of investigating the nature of compact objects as a data-driven question. GWTC-3 reported nearly 90 CBC candidates, while the released O4 catalogs~\cite{GWTC4,GWTC5} add nearly two hundred new candidates, bringing the number of reported detections to over $\sim300$. This growing sample has repeatedly revealed systems that challenge typical binary formation scenarios~\cite{LIGOScientific:2026ctl}, such as GW190814~\cite{GW190814} and GW230529\_181500~\cite{GW230529}, and very massive binaries, such as GW190521~\cite{GW190521} and GW231123\_135430~\cite{GW231123}; we hereafter refer to all events through their shortened GWYYMMDD notation, see the Appendix for their full names. Motivation for probing such outliers comes from the fact that compact objects are especially powerful probes of fundamental physics: they provide both a strong-field environment where departures from standard black hole (BH) and neutron star (NS) descriptions may become observationally relevant and, through GW emission, a direct messenger of those dynamics.

Studying compact binaries in the strong-field regime requires numerical relativity (NR), since the late inspiral, merger, and ringdown cannot be described reliably using weak-field or slow-motion approximations alone. Numerical simulations provide the fully nonlinear GR baseline against which small deviations must be identified, and they are essential for calibrating the phenomenological and effective-one-body waveform models used in GW parameter estimation~\cite{Ajith:2007qp,Ajith:2009bn,Khan:2015jqa,Husa:2015iqa,Hannam:2013oca,IMRPhenomXO4a,IMRPhenomXPHMSpinTaylor,IMRPhenomXPNR,MultipoleAsymmetry,Buonanno:1998gg,Buonanno:2000ef,Taracchini:2013rva,Buonanno:2005xu,SEOBNRv5PHM,SEOBNRv5PHMAsym}. This is especially important when testing compact object nature: a genuine non-BBH effect may appear only as a subtle change in the accumulated inspiral phase, the merger morphology, or the remnant ringdown, and must be distinguished from waveform systematics and GR effects such as spin precession, higher modes, eccentricity. Numerical studies of exotic compact object mergers, including boson-star binaries, have begun to show explicitly how non-BBH signals can either hide within the BBH waveform manifold with biased parameters or produce clear residual structure~\cite{Evstafyeva:2024qvp,Evstafyeva:2026juq}, underscoring the need for diagnostics that track deviations across the full inspiral-merger-ringdown (IMR) signal.

Current GW-based tests of the nature of compact objects primarily constrain deviations from the BBH hypothesis either through inspiral effects, such as modified phase evolution from tidal interactions and spin-induced multipolar structure, or through properties of the post-merger remnant such as the quasinormal mode (QNM) spectrum of  the ringdown~\cite{LIGOScientific:2021sio}. To the best of our knowledge, existing observational tests of exotic compact object (ECO) binaries have not yet directly targeted modifications to the merger morphology. As a first step in this direction Ref.~\cite{Ghosh:2025wex} introduced the ~\deco~model, which phenomenologically modifies the inspiral--merger transitions of a BBH waveform model (\!\!\bbh) through a single effective compactness parameter, $C_{\rm eff}$.

Ref.~\cite{Ghosh:2025wex} demonstrated that, in controlled injections with signal-to-noise ratio (SNR) $\sim 25$, the effective compactness parameter in~\deco~can be recovered with percent-level precision, and showed that the model can also be applied to real GW events. The natural next step is therefore to deploy this framework across a broad population of mergers and ask whether any events show evidence for non-BBH-like compactness. In this sense,~\deco~offers a new way of searching the catalog: not by targeting a specific exotic model, but by asking whether the merger morphology of any event departs systematically from the BBH expectation within this phenomenological parametrization. Applied at population scale, such an analysis opens up the possibility of identifying unusual mergers that would merit closer follow-up as candidate exotic binaries. It also reveals population-level features of the compactness distribution of astrophysical binaries, which we report in a companion \textit{Letter}~\cite{companion_letter}.

In addition to establishing the behaviour of~\deco~on observed BBH-like events, it is important to test whether the framework can recover genuinely non-BBH-like compactness in a realistic setting. We therefore complement the catalog analysis with binary neutron star (BNS) injections, which provide controlled signals with compactness values distinct from the BBH limit. These injections help us validate the ability of~\deco~in responding to departures from BBH merger morphology when present, and assess the limitations of such recovery in practice.

However, because the inferred compactness reflects a phenomenological deformation of the merger morphology, an apparent departure from the BBH expectation may arise not only from genuinely non-BBH structure, but also from waveform systematics, missing physics, or detector artefacts. Benchmarking the behaviour of~\deco~on real GW data is therefore essential before interpreting any merger as exotic. Motivated by this, we apply~\deco~to a subset of the mergers observed so far by the international GW detector network of LIGO-Virgo-Kagra (LVK)~\cite{LIGOScientific:2014pky,aLIGO:2020wna,Tse:2019wcy,VIRGO:2014yos,Virgo:2019juy,Virgo:2022ysc,Somiya:2011np,PhysRevD.88.043007,KAGRA:2020tym} and use the resulting population to establish an empirical baseline for BBH-like compactness posteriors. This baseline both reveals recurring posterior morphologies and systematic failure modes, and provides the context for the operational criteria defined in Sec.~\ref{sec:exotic_def} particularly the requirement that any apparent exclusion of the BBH value $C=0.5$ remain stable under robustness tests. In this way, the catalog analysis serves not only as a search for candidate exotic mergers, but also as the empirical context needed to interpret effective compactness measurements on real GW data using~\deco~.

We describe a binary using its total mass $M=m_1+m_2$, the mass ratio $q=m_2/m_1\leq 1$, and the two spin combinations most relevant for the waveform morphology. The effective aligned spin is denoted by $\chi_{\rm eff}$, while the spin precession parameter is denoted by $\chi_p$. In section~\ref{sec:setup} we place~\deco~in the context of complementary tests of exotic compact binaries and outline the analysis setup. Section~\ref{sec:trends} discusses the different types of effective compactness posterior we noted in the sample population, while Section~\ref{sec:results} presents the results across the sample, and a few special events from the fourth observing run (O4). Section~\ref{sec:residuals} shows how residual diagnostics can further inform searches for exotic binaries.

\section{Testing the nature of compact binaries}
\label{sec:setup}

Testing the nature of compact objects in GW observations involves, in principle, testing the nature of three objects: the two components of the binary and the remnant formed after merger. Existing tests broadly fall into two classes. The first targets the binary components during the inspiral, for example by allowing parametrized deviations in the waveform phase associated with finite-size effects such as spin-induced multipole moments~\cite{Krishnendu:2017shb,Krishnendu:2019tjp,krishnendu2025testingnaturecompactobjects}. The second targets the remnant, through BH spectroscopy with parametrized ringdown deviations~\cite{LIGOScientific:2021sio,Dreyer:2003bv,Berti:2005ys} or through searches for post-merger echoes~\cite{Uchikata:2019frs,Uchikata:2023mkw}. These complementary tests face distinct challenges: during the inspiral, the phase is the primary accurately measurable quantity sensitive to the nature of the components, but it is also affected by many other physical effects and possible modelling systematics; in the ringdown and post-merger regime, the available SNR is typically lower, and conclusions can be sensitive to assumptions such as the ringdown start time. 


The strength of the above approaches lies in the fact that their results can be interpreted in terms of specific physical effects tied either to the binary components or to the remnant. However, they probe only selected parts of the full coalescence signal. In particular, if deviations from the BBH signal are expected to become most prominent in the nonlinear merger regime, then purely inspiral-based or post-merger/ringdown-based tests may miss part of the relevant phenomenology. In addition, spin-induced multipole tests are most informative for binaries with appreciable component spins, which limits their applicability across the observed population. A full inspiral--merger--ringdown (IMR) test of ECO binaries, based on compactness as a generic property of compact objects, could therefore provide a useful probe of the strong-field merger regime while tracking deviations consistently across the full signal. A related class of tests is provided by IMR consistency tests (IMRCT)~\cite{Hughes:2004vw,Ghosh:2016qgn}, which check whether the weak-field inspiral and the strong-field merger--ringdown portions of the signal give mutually consistent source properties. These tests are powerful diagnostics of broad departures from the standard BBH description, including possible deviations from GR, non-BBH component structure, or other unmodelled physical effects. However, precisely because they are sensitive to many possible sources of inconsistency, their results are less directly interpretable as tests of compact object nature alone.

\subsection{Compactness}
\label{sec:compactness}

Compactness is conventionally defined as $C\coloneqq m/r$ in geometrized units, for an object of mass $m$ and radius $r$. For a binary, one can analogously introduce a geometric compactness proxy by comparing the total mass to a characteristic length scale of the system close to merger. Taking this length scale to be set by the sum of the characteristic radii of the two objects gives
\begin{equation}
C_{\rm geometric} = \frac{m_A + m_B}{r_A+ r_B},
\label{eq:C_geometric}
\end{equation}
where $m_i$ and $r_i$ denote the mass and characteristic radius of object ($i=A,B$), respectively. Noting that for a Kerr black hole in Boyer-Lindquist coordinates,
\be
r_i\equiv r_i^{\rm horizon} = m_i+\sqrt{m_i^2-a_i^2} =m_i \left( 1+\sqrt{1-\chi_i^2} \right), 
\ee
where $a_i$ is the Kerr parameter $(|a_i|\leq m_i)$ and $\chi_i =\frac{a_i}{m_i}$, is the dimensionless spin, we can rewrite Eq.~\eqref{eq:C_geometric} as,
\begin{equation}
C_{\rm  geometric} = \frac{m_A + m_B}{m_A \left(1+\sqrt{1-\chi_A^2}\right)+ m_B\left(1+\sqrt{1-\chi_B^2}\right)}\;.
\end{equation}
Written this way, it is easy to see $C_{\rm geometric} \geq0.5$ for astrophysical BHs with any combination of component masses and $|\chi_i|\leq1$. For example, for $\chi_A = \chi_B=\chi$ it simplifies to $1/(1+\sqrt{1-\chi^2})$, yielding $C_{\rm geometric} > 0.5$ for non-zero $\chi$. However there is no unique, gauge-independent interpretation of this quantity as a binary compactness near merger, and it is not directly observable in GW data. We therefore use $C_{\rm eff}$ which tracks the waveform amplitude morphology, with ($C_{\rm eff}=0.5$) chosen by convention to reproduce the baseline BBH model.

\subsection{\deco: A Brief Recap}
\label{sec:deco}

Lower compactness for the individual components imply larger objects and therefore an earlier onset of the post-contact stage of the signal. ~\deco~is a modification of a standard aligned-spin BBH waveform model 
that adds an extra degree of freedom to\bbh~\cite{Husa:2015iqa,Khan:2015jqa}, by taking the impact of lower compactness into account. Specifically,~\deco~allows the inspiral to terminate earlier than in the BBH case, before smoothly transitioning to a BBH-like continuation beyond that point. This continuation is not intended as a physical model of ECO post-contact dynamics,  but as a phenomenological prescription that preserves continuity with the BBH limit. This choice reflects a broader limitation in the current literature: fully relativistic simulations of ECO mergers remain scarce, with boson-star binaries being one of the few cases studied in detail~\cite{Palenzuela:2006wp,Palenzuela:2007dm,Dietrich:2018bvi,Clough:2018exo,Evstafyeva:2024qvp,Siemonsen:2023hko}. 
While~\deco~can be generalized to include finite-size effects in the phase and post-contact/ringdown deviations, in its present form we focus on finite-size effects encoded only in the waveform amplitude morphology near merger.

\begin{figure}[h]
	\includegraphics[width=0.45\textwidth]{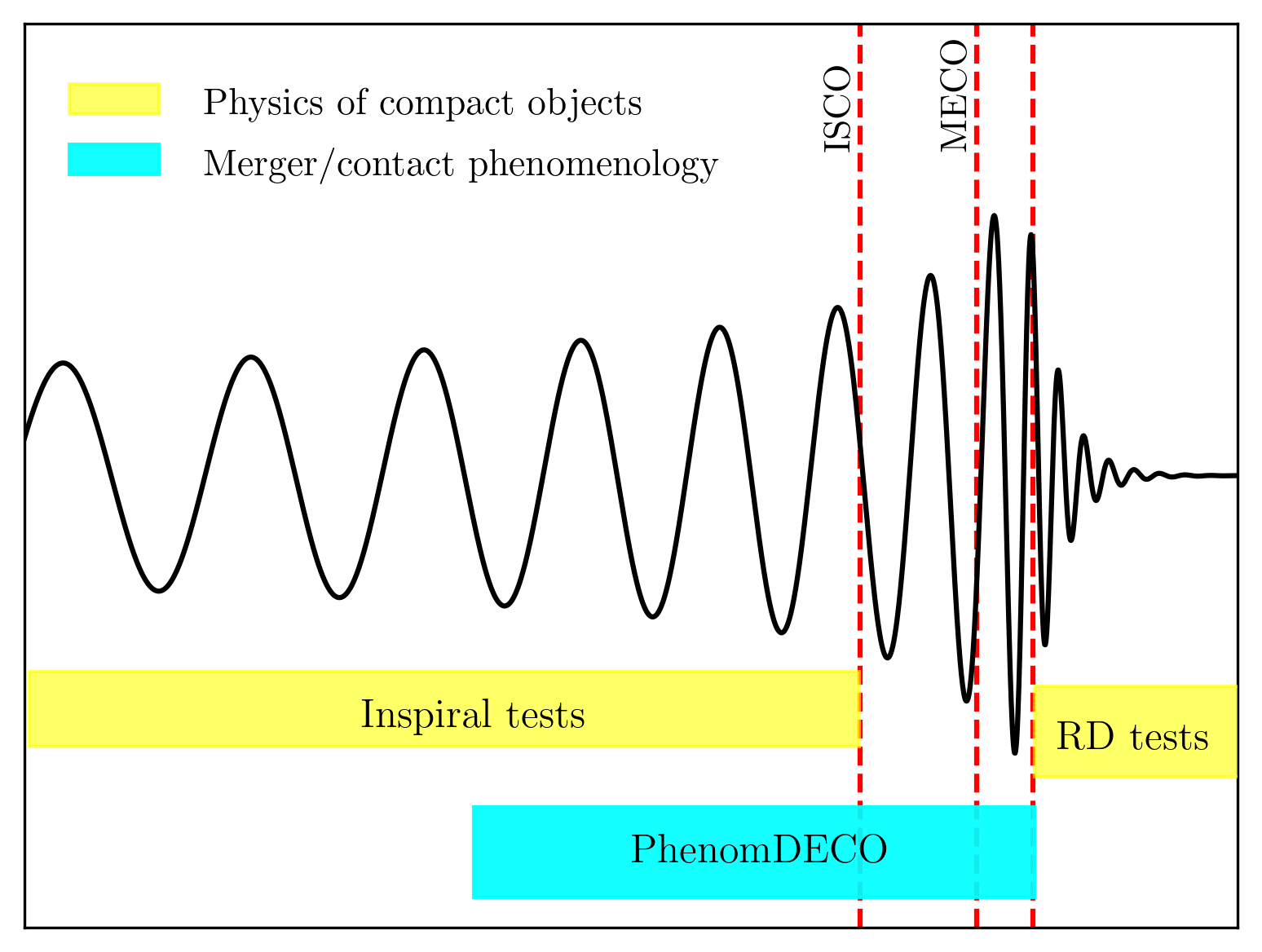}
	\caption{The time domain GW signal associated with a BBH merger, with the binary mass scaled out, is shown in black; the characteristic dimensionless frequencies are indicated by dashed vertical red lines. The tests that probe the physics of individual BHs are denoted in yellow (ringdown abbreviated as RD), while the test of merger phenomenology is indicated in cyan.}
	\label{tgr_context}
\end{figure}

The amplitude morphology is modified through $C_{\rm eff}$ that controls the transition from inspiral to merger--ringdown according to how close the objects can approach at merger. While this parameter is motivated by the same geometric intuition as $C_{\rm geometric}$, the two quantities are not interchangeable. This is because~\deco~does not currently model spin dependence of the compactness parametrization self-consistently. Specifically, the present parametrization maps the inspiral--merger transition frequencies of all BBHs, independent of component masses or spins, to the geometric compactness of a nonspinning BBH, i.e., 0.5. By construction~\bbh~already incorporates component spins aligned with the orbital angular momentum through a mass-weighted spin combination~\cite{Ajith:2009bn}, but spin-dependent changes in $C_{\rm geometric}$ are not mapped into the current parametrization. Including an explicit spin dependence in the compactness mapping may therefore improve the physical interpretation of inferred $C_{\rm eff}$. Nevertheless, the central conclusion remains unchanged: within this parametrization, BBH-like configurations should not correspond to $C_{\rm eff}<0.5$. ~\deco~should therefore be interpreted as a microphysics-agnostic probe of whether the merger morphology appears BBH-like, and only as an approximate measure of the objects' compactness. In particular, a robust measure of $C_{\rm eff}<0.5$ does tell us that at least one of the objects is not a BH.


Figure~\ref{tgr_context} illustrates the stage of the binary coalescence at which~\deco~is most sensitive to departures from a standard BBH signal. Three characteristic frequencies are indicated in red: the orbital frequency at the innermost stable circular orbit (ISCO) of a non-spinning black hole with mass equal to the total binary mass, which is often used as a rough upper limit for the regime in which the post-Newtonian (PN) approximation remains reliable; the minimum-energy circular orbit (MECO) frequency, often a better proxy for the inspiral-to-plunge transition, especially for comparable-mass binaries; finally, the characteristic ringdown frequency of the remnant. As denoted on the figure, the inspiral-based tests of nature of components rely on the dephasing before ISCO, while ringdown-based tests track departures from a Kerr geometry of the remnant.~\deco~on the other hand probes the merger phenomenology through late-inspiral and merger/contact. It is complementary to existing tests in the part of the IMR signal it probes and allows us to test the nature of compact objects from a different angle: how BBH-like is the merger morphology of a GW observation? This immediately implies: (i)~\deco~would be generally more informative on deviations from BBH for shorter signals (i.e., higher-mass binaries), for which a larger fraction of the observed SNR lies near merger; and (ii) it can serve as a complementary cross-check to IMRCT, probing for tensions between the weak-field inspiral dynamics and the late-inspiral/merger waveform morphology, since it matches the corresponding BBH approximant exactly in the inspiral regime. 

\subsection{\deco~waveform systematics}
\label{sec:systematics}

\begin{figure*}[!t]
	\includegraphics[width=0.86\textwidth]{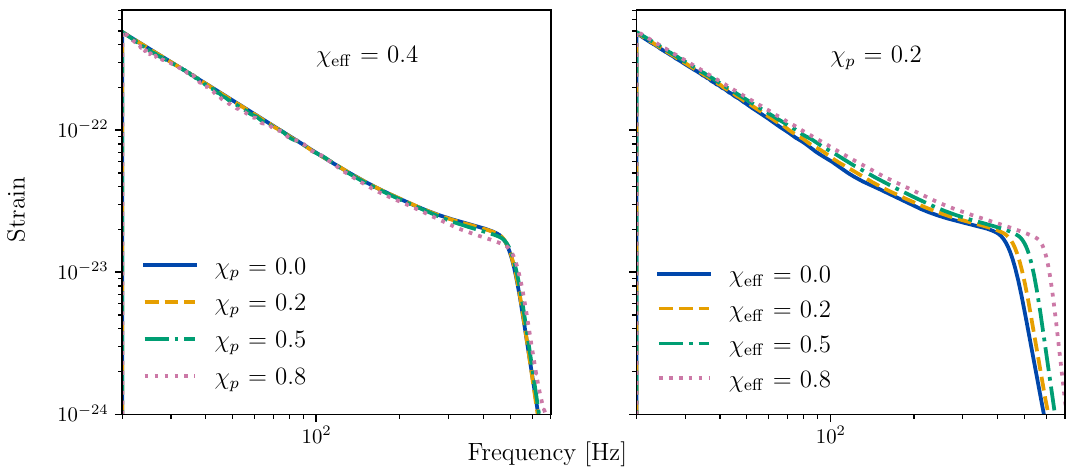}
	\caption{Frequency-domain amplitude for precessing binaries generated with the standard precessing binary waveform model~\xphm~for an equal-mass binary with $M=40~\msun$. The left panel shows waveforms with fixed $\chi_{\rm eff}=0.4$ and increasing precessing spin $\chi_p=\{0.0,0.2,0.5,0.8\}$, shown with solid, dashed, dash-dotted, and dotted lines, respectively. The right panel shows waveforms with fixed $\chi_p=0.2$ and increasing effective aligned spin $\chi_{\rm eff}=\{0.0,0.2,0.5,0.8\}$, shown with solid, dashed, dash-dotted, and dotted lines, respectively.}
	\label{fig:prec_waveforms}
\end{figure*}

The baseline BBH model used in~\deco~does not include higher multipoles or precessional effects from in-plane spins. These missing effects can, in principle, introduce waveform-modelling systematics in effective compactness measurements. We distinguish here between two possible effects: a direct degeneracy, in which precession produces an amplitude signature similar to the compactness-dependent modification targeted by~\deco, and an indirect bias, in which precession biases the recovery of other BBH parameters that are correlated with compactness. The first possibility appears to be limited. This is illustrated using the frequency-domain amplitude of the standard precessing BBH waveform model~\xphm~\cite{Pratten:2020ceb}. As shown in the left panel of Fig.~\ref{fig:prec_waveforms}, the amplitude near merger exhibits very little sensitivity to precessing spins. This contrasts with the right panel, where aligned spin has a little more noticeable impact on the amplitude morphology. We therefore expect direct degeneracies between precession and effective compactness measurement to be limited. 

\begin{figure*}[!t]
	\includegraphics[width=0.9\textwidth]{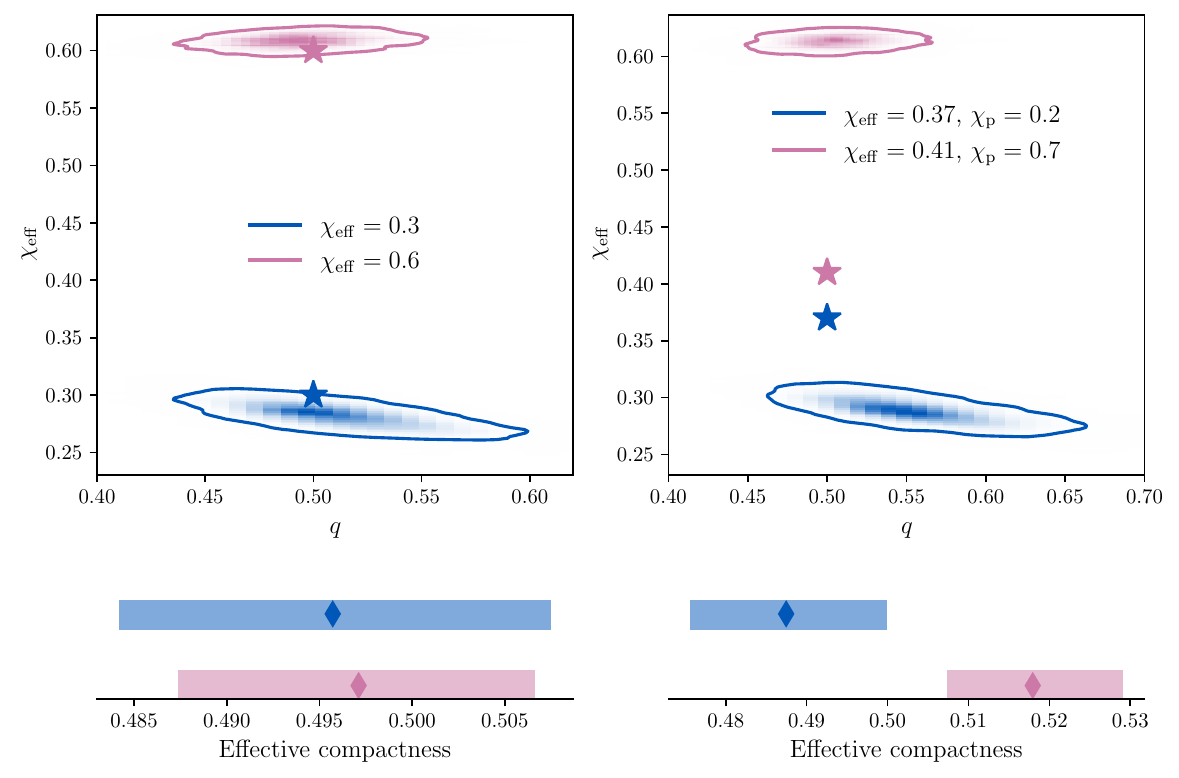}
	\caption{Two-dimensional posterior probability distributions in mass ratio $q$ and effective aligned spin $\chi_{\rm eff}$ for unequal-mass BBH injections with $M=45~\msun$ and $q=0.5$. The left panel shows aligned-spin injections at ${\rm SNR}\sim50$, with $\chi_{\rm eff}=0.3$ shown in blue and $\chi_{\rm eff}=0.6$ shown in pink. The right panel shows precessing injections at ${\rm SNR}\sim12$, with $(\chi_{\rm eff},\chi_p)=(0.37,0.2)$ shown in blue and $(\chi_{\rm eff},\chi_p)=(0.41,0.7)$ shown in pink. In both panels, the injected values are marked by stars of corresponding colour. The lower subpanels show the corresponding effective compactness posteriors as horizontal box plots, with the marker indicating the median and the box edges enclosing the $90\%$ CI. The precessing injections were performed at an orbital inclination $\theta_{\rm{jn}} = 0.4$, to ensure precessional modulations were not suppressed.
	}
\label{fig:systematics}
\end{figure*}

We test this expectation with unequal-mass BBH injections with $q=0.5$ and $M=45~\msun$, generated with~\xphm~and recovered with~\deco~in a Bayesian analysis assuming advanced LIGO detector sensitivity~\cite{LIGOScientific:2014pky}. As a control case, we first consider non-precessing injections with $\chi_{\rm eff}=\{0.3,0.6\}$ at ${\rm SNR}\sim50$, shown in the left panel of Fig.~\ref{fig:systematics}. In these cases all parameters are well recovered, including the mass ratio, $\chi_{\rm eff}$, and effective compactness, with only mild degeneracies between $\chi_{\rm eff}$ and effective compactness. The precessing injections, shown in the right panel of Fig.~\ref{fig:systematics}, illustrate the second effect; we inject signals with $\chi_p=\{0.2,0.7\}$, moderate $\chi_{\rm eff}$, and ${\rm SNR}\sim12$. In these cases,~\deco~recovers the signals with mild to moderate biases in $\chi_{\rm{eff}}$ and larger departures from $C_{\rm eff}=0.5$; the bias persists even when we restrict to posterior samples with $C_{\rm eff}=0.5$, suggesting that the bias is not introduced by the compactness degree of freedom but instead reflects the recovery of precessing signals with an aligned-spin model.
Given the mild correlation between $\chi_{\rm eff}$ and effective compactness seen in the non-precessing injections, this suggests that the compactness bias may be driven, at least in part, by the biased recovery of $\chi_{\rm eff}$. However, these injections indicate that none of these systematics can mimic the low-compactness measure that we would classify as non-BBH. A non-BBH classification by~\deco~is therefore less directly exposed to the systematics that affect tests based on accumulated phase differences, such as dephasing caused by inaccuracies in, or omissions from, the modelling of various general relativistic (GR) effects. We discuss the systematic biases introduced by neglecting tidal effects in sec.~\ref{sec:bns_hybrid}.

\begin{figure*}[!t]
	\includegraphics[width=0.91\textwidth]{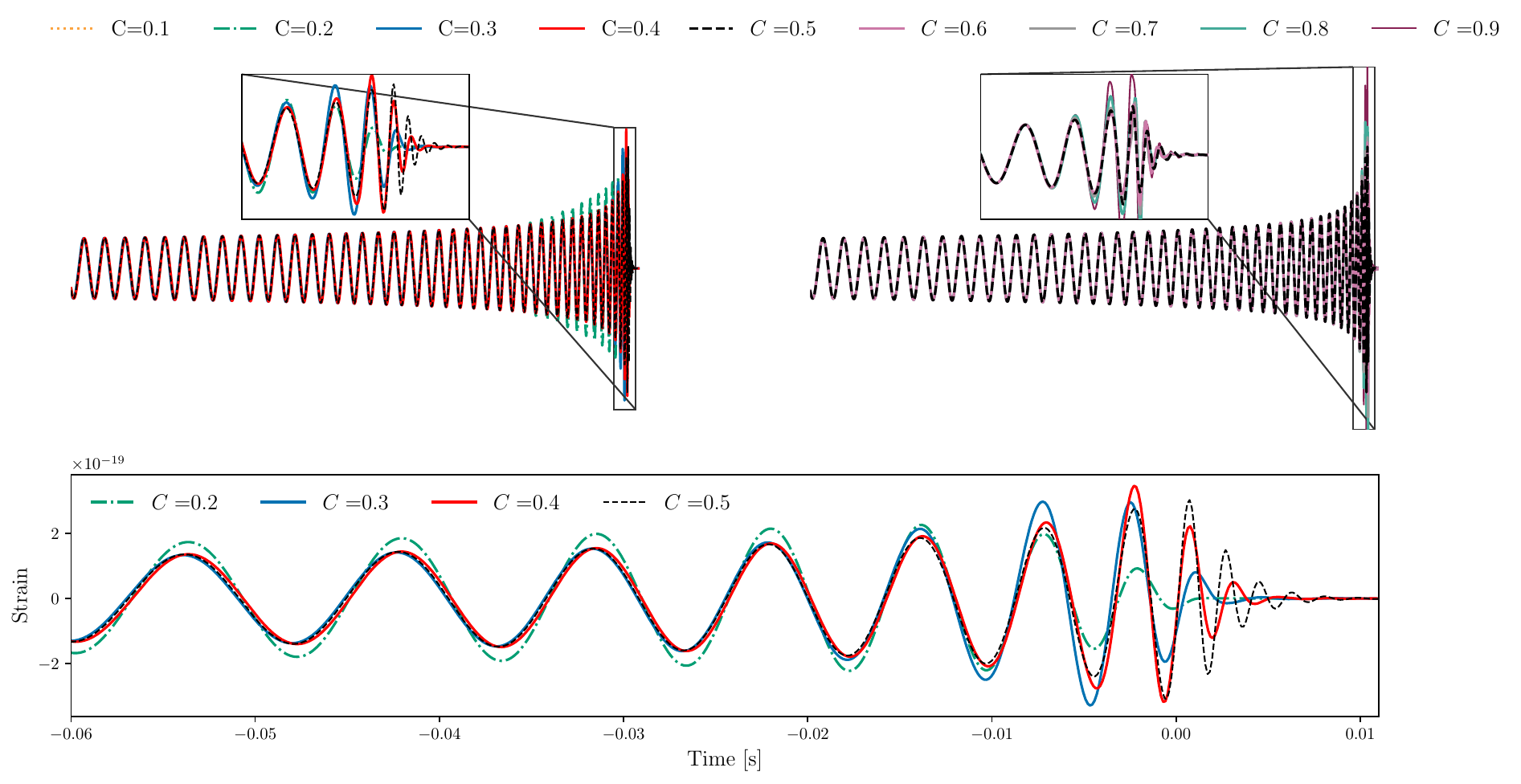}
	\caption{Time-domain~\deco~waveforms for different values of the effective compactness $C_{\rm eff}=[0.1,0.2,0.3,0.4,0.6,0.7,0.8,0.9]$, for a fixed mass ratio of $q=0.5$ and $M=30 \msun$. The top left panel shows waveforms with $C_{\rm eff}<0.5$, with the BBH expectation $C_{\rm eff}=0.5$ shown as a black dashed line. The top right panel shows that cases with $C_{\rm eff}>0.5$ are largely identical to the BBH waveform. The lower panel shows the last few cycles where the early end of inspiral based on compactness is easy to note. Signals with $C_{\rm eff}<0.2$ are significantly shorter and terminate earlier than the time window shown.
}
\label{fig:td_wfs}
\end{figure*}

The most prominent systematic we find with~\deco~is a degeneracy in the waveform for $C_{\rm eff}\geq0.5$. This is demonstrated in Fig.~\ref{fig:td_wfs}, where we show the time-domain behaviour of~\deco~for different values of the effective compactness. Since all waveforms share the same orbital parameters and differ only in $C_{\rm eff}$, we impose a physically motivated alignment by requiring them to reach the starting frequency of $20$ Hz at the same time. With this alignment, the waveforms for $C_{\rm eff}>0.5$ remain largely indistinguishable from the BBH case across the full inspiral--merger--ringdown signal, as shown in the top-right panel (see also Fig.~2 in Ref.~\cite{companion_letter}). By contrast, the $C_{\rm eff}<0.5$ waveforms cannot be aligned with the BBH waveform throughout the full signal. Instead, they exhibit the expected behaviour: the transition from inspiral to merger--ringdown occurs earlier and the waveform is shorter, as illustrated by the lower panel of Fig.~\ref{fig:td_wfs}. The top left panel shows that these waveforms can be aligned in early inspiral. We therefore interpret support for $C_{\rm eff}>0.5$ to be consistent with BBH within this parametrization. 


\subsection{Impact of missing tidal effects on effective compactness inference}
\label{sec:bns_hybrid}

Since~\deco~is constructed as a phenomenological modification of a BBH baseline waveform, an important expected systematic in the recovery of $C_{\rm eff}$ is the absence of tidal effects. Such effects are generically expected for neutron stars and may also arise for other non-BBH compact objects, and can therefore bias the interpretation of the compactness inferred hrough Bayesian inference. We use BNS injections as a controlled stress test of this systematic. To ensure that the merger, where the~\deco~modification is most directly probed, remains visible in band, we perform these analyses using the Einstein Telescope sensitivity curve~\cite{Hild:2010id}. To sample the posterior distribution, we use {\texttt{dynesty}}~\cite{Speagle:2019ivv} via {\texttt{bilby}}~\cite{bilby,Romero-Shaw:2020owr}, as has been done by the LVK consistently since the third GW observing run. To interface with~\deco~we use the modifications described in Ref.~\cite{Ghosh:2025wex}. We also assume wide and agnostic priors following the conventions adopted by the LVK, \emph{e.g.} we sample uniformly in component masses, and we use 2000 live points to improve sampler convergence.

We first analyze the BNS NR waveform BAM-0094 from the CoRe database~\cite{Dietrich:2018phi}, which contains only $\sim 12$ cycles before merger. As discussed in the \textit{Letter}~\cite{companion_letter}, this provides a clear validation case with~\deco~recovering a low effective compactness, $C_{\rm eff}\simeq 0.15$, and confidently excludes the BBH-like value $C_{\rm eff}=0.5$. The inferred compactness also agrees well with the approximate geometric estimate $C_{\rm geometric}$. However, the same analysis shows significant biases in the recovered mass ratio and $\chi_{\rm eff}$, together with smaller biases in the chirp mass. To test whether these biases arise from the BBH baseline waveform rather than the compactness-based modification in~\deco~,we analyse a hybrid BNS waveform from the CoRe database~\cite{Dietrich:2018phi}, constructed by matching analytical inspiral predictions to the late-inspiral and merger portion of the NR simulation~\cite{Dudi:2018jzn}. Although the hybridisation significantly increases the inspiral of the NR waveform, the starting frequency is $\sim 200\, \mathrm{Hz}$. We therefore only analyse data between $200 < f < 1680\, \mathrm{Hz}$. We find that the mass-ratio bias persists in the hybrid analysis, consistent with the findings of Ref.~\cite{Dudi:2018jzn}. This supports the interpretation that the bias is driven by the baseline waveform model~\bbh.

\begin{figure}[h]
	\includegraphics[width=0.46\textwidth]{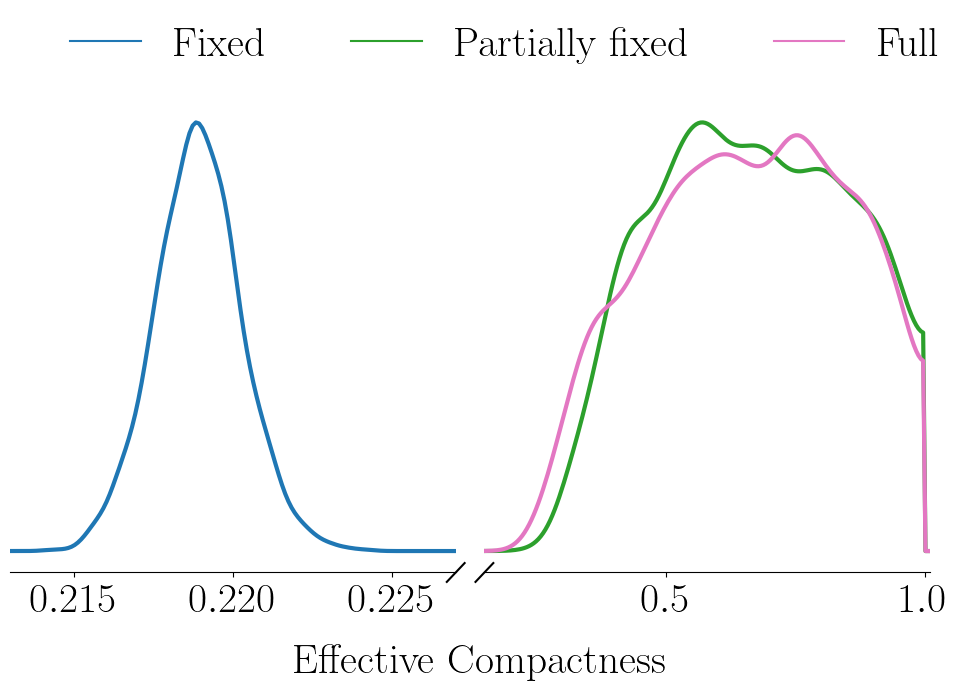}
	\caption{The inferred effective compactness when analyzing an NR-hybrid waveform injected in ET. We show the posterior distribution obtained when using the full likelihood (pink), when fixing the chirp mass, total mass and mass ratio to their true values (green), and when fixing all parameters except for the luminosity distance, effective compactness and merger time (blue).}
	\label{fig:BNS_hybrid_compactness}
\end{figure}

The hybrid waveform analysis showed similar biases in the mass ratio, although the bias in $\chi_{\rm{eff}}$ was reduced. However, unlike the NR case, the compactness posterior for the hybrid waveform prefers a BBH-like value. This is unexpected because the merger remains in band for both injections, and one might therefore expect the compactness information near merger to be recovered in both cases. We performed several diagnostic tests, including truncating the post-merger signal to one cycle after the peak, reducing the number of inspiral cycles (from $3000$ cycles in the original hybrid signal), and restricting the set of parameters the likelihood samples over. A non-BBH compactness was recovered only in the restricted analysis where the likelihood sampled over just $C_{\rm eff}$, geocentric trigger time, and luminosity distance while fixing all other intrinsic and extrinsic parameters, as shown in Fig.~\ref{fig:BNS_hybrid_compactness}. Furthermore following a prescription similar to~\cite{companion_letter}, we estimate a $C_{\rm geometric}\simeq 0.15$ which is inconsistent with the inferred value even in the restricted likelihood analysis. This indicates that although its performance remains robust for transient signals, as demonstrated by the BNS NR injection discussed above, a prescription for tidal effects will be necessary before~\deco~can be reliably applied to long-duration signals.

\section{Broad trends and non-BBH classification}
\label{sec:trends}
We perform Bayesian inference with~\deco~on all high-significance BBH events from GWTC-3, defined by a false-alarm rate < $0.25 \, \rm{yr}^{-1}$, for a total of 69 events  listed in Table~\ref{tab:event_catalog_summary} in Appendix~\ref{app:appendix1}. As in Sec.~\ref{sec:bns_hybrid}, all Bayesian analyses are performed with {\texttt{dynesty}} via {\texttt{bilby}}. Throughout this and subsequent sections, we use the same priors and sampler settings as those done in the original LVK analysis unless otherwise stated. The only difference is that we project the priors into the non-precessing space, i.e. we marginalize over the in-plane spin directions, since~\deco~does not incorporate precession effects, and we use a uniform prior for the effective compactness over the interval[0.1,0.99].

\subsection{Trends in effective compactness posteriors}
In this section, we summarize the trends observed in effective compactness posterior proability distributions from Bayesian inference across the events quoted above. Under the approximation adopted in~\deco, $C_{\rm eff}=0.5$ represents the reference BBH limit. BBH-like events are therefore expected to remain consistent with this value, although waveform systematics and noise fluctuations can mildly displace the peak below 0.5 or introduce additional support at $C_{\rm eff}\geq0.5$, as discussed in Sec.~\ref{sec:systematics}.
The studies in Sec.~\ref{sec:setup} suggest that precession can produce small downward shifts in the inferred compactness for BBH signals, even in the absence of noise. However, the observed shifts are not large enough to account for the low-compactness departures considered here, such as $C_{\rm eff}\leq0.4$. This underscores the scope of~\deco~: \textit{it is best suited to identify candidate ECO binaries with effective compactness significantly below the BBH limit}. We find three broad trends across the full set of events that we discuss below. 
\subsubsection{Dominant peak near C=0.5}
\begin{figure*}[!t]
	\includegraphics[width=0.98\textwidth]{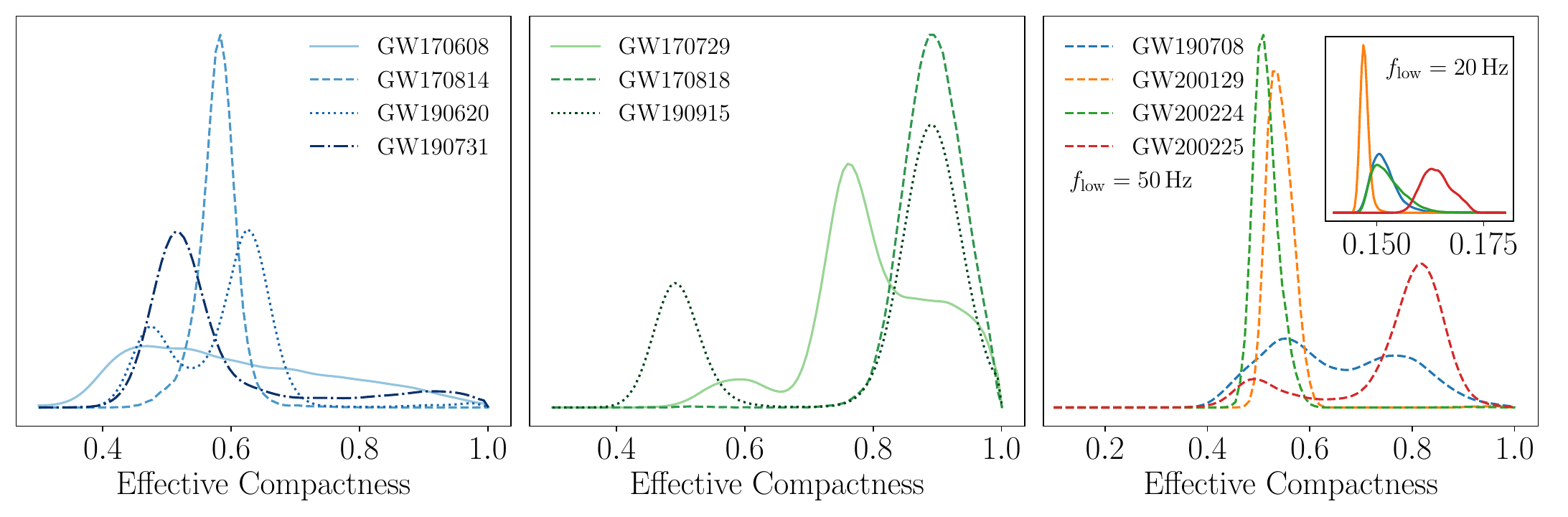}
	\caption{Plot showing the inferred compactness for events that show \textit{Left:} typical behaviour (defined in text); \textit{Middle:} significant support above $C>0.5$; \textit{Right:} a clear preference for $C\sim0.15$ when analysing data with frequencies $> 20\, \mathrm{Hz}$, but negligible support for $C < 0.5$ when analysing data with frequencies $> 50\, \mathrm{Hz}$ in Livingston. In the \textit{Right} panel, we explicitly show the posterior distribution obtained when consistently using a low-frequency cutoff of $20\, \mathrm{Hz}$ in all detectors in the inset, and the dashed lines show the posteriors obtained when analyzing Livingston data above 50 Hz and Hanford (and Virgo, where available) from the usual $20$ Hz.}
\label{fig:different_C_dist}
\end{figure*}
About two thirds of the events show posteriors consistent with BBH, which we classify as typical behaviour. As shown in Figure~\ref{fig:different_C_dist}, this category includes posteriors that: (i) exhibit a clear peak near $C=0.5$ (GW190731); (ii) peak at 0.5, but with a very broad 90\% credible interval (CI) (GW170608); (iii) are bimodal, with peaks displaced from $C=0.5$ but including it well within the 90\% CI (GW190620); and (iv) have a lower 90\% CI bound only marginally above 0.5 (GW170814) as this kind of behaviour can happen due to missing physics (cf. Fig.~\ref{fig:systematics}). For these events, the inference on all other intrinsic parameters were found to be consistent with that reported in GWTC-3.
\subsubsection{Support for C> 0.5}
We found significant support for a compactness much higher than 0.5 for about ten events. Only one event shows a clear preference for $C\sim0.9$ (GW170818), and in two other events $C=0.5$ is excluded at $90\%$ lower CI. In all the other cases, although the posterior probability favours a higher compactness, $C= 0.5$ was always included in the 90\% CI, as shown in Fig.~\ref{fig:different_C_dist}. We do not observe biases in the other inferred parameters for these events, nor do we find correlations with quantities such as $q$ or $\chi_{\rm{eff}}$ that would explain the preference for higher compactness. Likewise, neither the total mass nor the SNR of these events provides additional insight into their anomalous behaviour. Given these observations, together with the discussion on systematics of effective compactness in Sec.~\ref{sec:setup}, we interpret this behaviour as BBH-like rather than as evidence for exotic physics.

\label{sec:lowc}
\begin{figure}[!t]
	\includegraphics[width=0.44\textwidth]{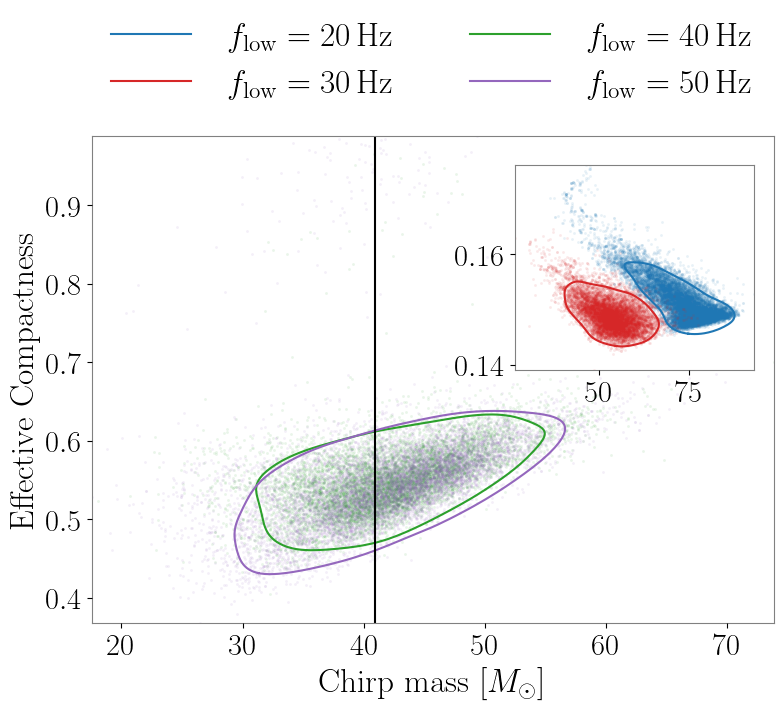}
	\caption{Plot showing the 2D marginalized posterior distribution for the chirp-mass and effective compactness when analysing GW200224 with progressively higher low-frequency cutoffs in the Livingston detector, $f_{\rm low} = [20, 30, 40, 50]~\mathrm{Hz}$; we consistently used $f_{\rm low} = 20\, \mathrm{Hz}$ in the Hanford detector. The solid black vertical line shows the median chirp-mass inferred with~\xphm~ by the LVK, and we obtain consistent results with $f_{\rm low} = 40~\mathrm{Hz}$. This is also the cutoff frequency at which the compactness posterior becomes consistent with BBH-like behavior. The inset shows a zoomed in region of the parameter space, focusing on $C\sim 0.15$.}
\label{fig:flow_dep}
\end{figure}

\subsubsection{Spuriously low measure of compactness}

Nominally, a compactness posterior whose upper 90\% CI lies below 0.5 would suggest that one or both components are of exotic origin. However, as discussed above, modest downward deviations from 0.5 can arise for a variety of non-exotic reasons. We therefore regard a compactness posterior as potentially suggestive of non-BBH structure only when both the posterior peak and the upper edge of the 90\% CI lie below $C\simeq0.4$. We note that none of the events in our study exhibited this behaviour.

The events considered in this subsection fall into a different category. In the initial analyses, roughly a dozen events showed a sharp low-compactness (low-C) mode near $C\sim 0.14-0.17$, excluding the BBH value at $90\%$ upper CI. However, these runs failed to identify the signal at the known geocentric coalescence time, accompanied by inconsistent chirp mass recovery and biases in other parameters. These pathological signatures are sufficiently pronounced to make the corresponding compactness measurements unreliable. The low-C modes are therefore best interpreted as arising from an interplay between the sampler and a short-duration burst of power near the signal, rather than as evidence for anomalously low compactness in the astrophysical source.

The events do not cluster in any particular region of the $\mathcal{M}$-$\chi_{\mathrm{eff}}$ parameter space, nor did we find any clear trend with their SNR, suggesting that the failure mode is not simply associated with intrinsic source parameters or low-SNR events. The only event in this set known to be associated with data-quality issues is the precessing BBH GW200129~\cite{Hannam:2021pit,Payne:2022spz}. However, we verified that this issue is unlikely to explain the anomalous compactness recovery, since the coalescence time preferred by~\deco~lies approximately 2s after the known coalescence time.

Another feature emerged from analyses of individual-detector data restricted to different frequency bands. Across a series of tests, we found that the anomalous low-C inference were primarily driven by the low-frequency content of the Livingston data. We note that low-frequency scattered-light noise in Livingston during O3, particularly in the 10–40Hz band, has also recently been discussed in Ref.~\cite{nandi2026scatteredlightnoiseligo}. While we do not attempt to identify a specific noise artefact in each event, this provides useful context for why the Livingston low-frequency data may be especially susceptible to this failure mode.

Figure~\ref{fig:flow_dep} shows that for GW200224~\cite{GWTC3}, the inferred effective compactness varies significantly as a function of frequency. If the low-C was truly astrophysical, we would expect the compactness measurement to be independent of the frequency content included. Specifically, when using $f_{\rm low} \geq 40~\mathrm{Hz}$ in the Livingston detector, we obtain a BBH like compactness. We also recover a chirp mass distribution comparable to that in the LVK analysis with~\xphm. Repeating this analysis for the other affected events showed that the cutoff required to recover the correct signal time varied from event to event. Nevertheless, using $f_{\rm low}=50$ Hz for the Livingston data alone removed the low-C peak in all cases and restored typical behaviour, as shown in the right panel of Fig.~\ref{fig:different_C_dist}. This strongly refutes an astrophysical origin for these anomalies; we therefore use posterior samples from the analysis that uses Livingston data above $50$ Hz for the rest of the paper.

\subsection{Exotic binary characterisation with~\deco~}
\label{sec:exotic_def}
We define here the criteria for a binary to be labelled exotic from the results of Bayesian inference with~\deco. These criteria are constructed to reflect the scope of the model, potential systematic biases in compactness inference arising from the omission of known physical effects, and susceptibility to noise artefacts:

\begin{itemize}

\item multidetector event
\item $M>10 \msun$ to ensure that the transition to merger lies within the sensitivity of current LVK detectors;
\item the effective compactness posterior excludes the BBH value $C=0.5$ from its 90\% CI, with the posterior support lying predominantly at $C<0.5$;
\item this exclusion of $C=0.5$ persists under robustness tests in which the lower analysis frequency is varied over $f_{\rm low}\in[20,50]\,{\rm Hz}$ in each detector;
\item cases in which the posterior excludes $C=0.5$ only in favour of $C>0.5$ are not classified as exotic, since $C>0.5$ is interpreted as BBH-like behaviour within the present parametrization rather than as a physical compactness.

\end{itemize}

It is worth noting that the choice of the frequency range for criterion 4 is heuristic, based on current understanding of data quality. As we discuss in section~\ref{sec:gw231123} this condition may need to be revisited as our understanding of low-frequency detector artefacts and data quality evolves. 

\section{Results}
\label{sec:results}
\begin{figure}[!t]
	\includegraphics[width=0.48\textwidth]{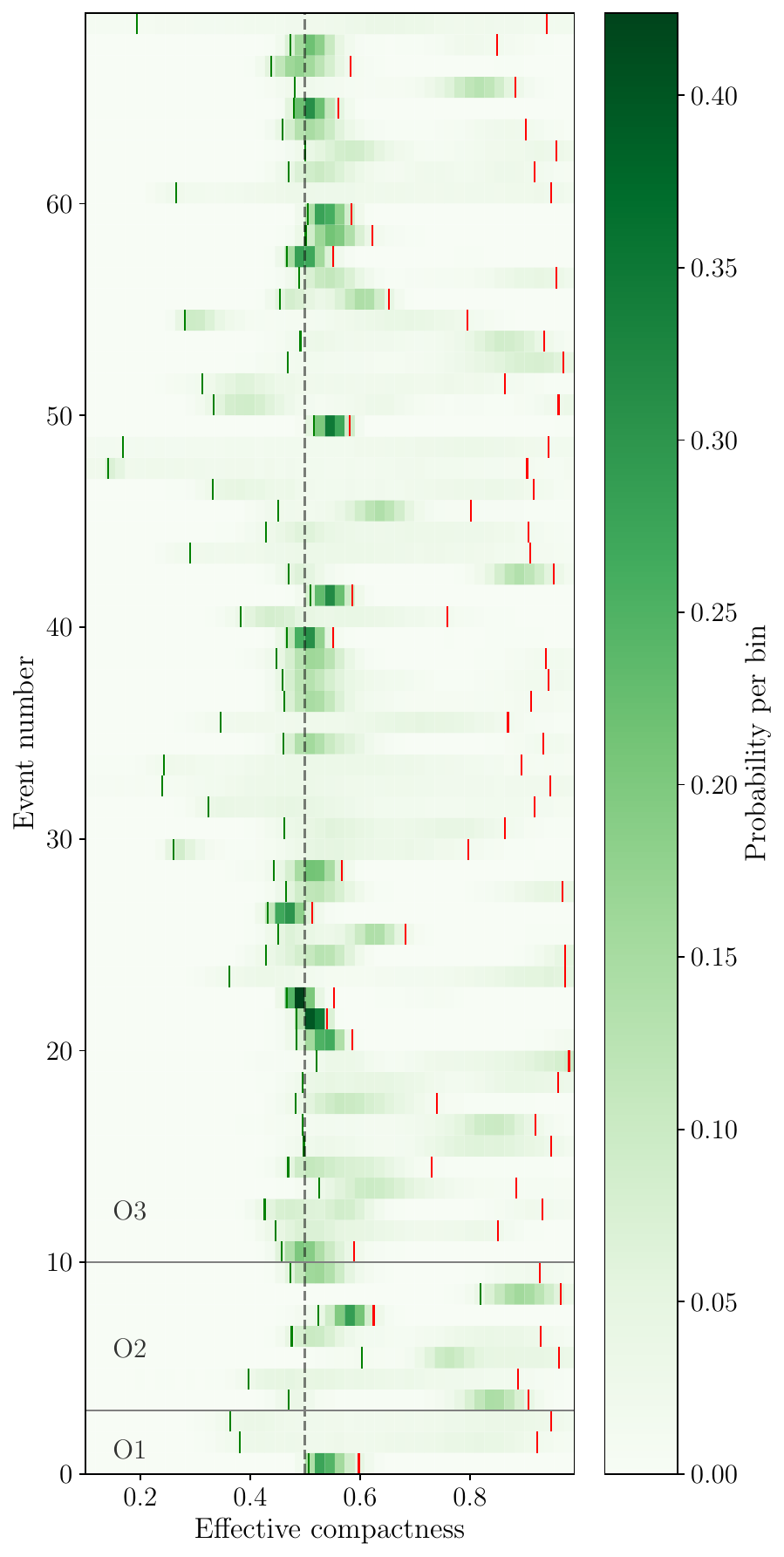}
	\caption{Compactness posterior distributions for the analyzed GWTC-3 events, shown as normalized probability per bin on a common set of effective compactness bins. Each row corresponds to one event, indexed as listed in Table~\ref{tab:event_catalog_summary} of Appendix~\ref{app:appendix1}, and the color indicates the normalized posterior probability in each bin. Darker regions indicate more sharply peaked compactness support, while lighter regions indicate broader or less localized posteriors. The grey vertical dashed line marks the BBH reference value $C_{\rm eff}=0.5$. For each event, the lower and upper bounds of the $90\%$ CI are indicated by green and red vertical bars, respectively. Events that initially showed a sharp low-C mode near $C_{\rm eff}\simeq0.15$ are shown here after reanalysing with a low-frequency cutoff of $50$ Hz in Livingston data; in all the cases, the low-C support is removed and the posterior becomes consistent with BBH-like compactness. Notably, since none of the upper $90\%$ CI lie below $C_{\rm eff}=0.5$, all analyzed events remain consistent with a BBH classification under our criteria.}
\label{fig:full}
\end{figure}

Figure~\ref{fig:full} summarizes the effective compactness posteriors for the full set of analyzed GWTC-3 events. The overall distribution shows a preference for BBH-like compactness values close to $C\simeq0.5$, and no event satisfies the non-BBH classification criteria defined in Sec.~\ref{sec:exotic_def}. We note a general preference for compactness close to 0.5. However, a robust population-level constraint requires hierarchical inference, which we present elsewhere~\cite{companion_letter}. In this section, we discuss the effective compactness estimates for several notable events. We first consider unequal-mass binaries, for which shorter merger morphology of a low-compactness signal could in principle be partially mimicked by biases in the total mass or mass ratio. We also examine a couple of notable O4 events like GW250114~\cite{LIGOScientific:2025rid}, selected for of its high SNR, and GW231123 due to its since its unusually high mass.
We discuss GW231123~\cite{GW231123} separately, as its short in-band duration, and challenging waveform systematics make it a natural target for exploring alternatives to a standard BBH interpretation.

\subsection{Unequal mass binaries}
\label{sec:mass_asymmetric}

Reference~\cite{Ghosh:2025wex} showed that, when ECO signals are characterized by an earlier end to the inspiral, they can be recovered by BBH templates with more asymmetric mass ratios. Events with large inferred mass ratios are therefore most likely candidates for mischaracterisation as ECO binaries within this framework. GW190412~\cite{LIGOScientific:2020stg} and GW190814 
are significantly unequal in masses with $q\sim3.5$ and $q\sim9$, respectively and therefore plausible candidates for such mischaracterisation.  However, with~\deco~ we find them both to be consistent with BBH: GW190814 shows a single peak at $C\sim0.53$, while the compactness posterior is bimodal and broad for GW190412, with a secondary peak at $C\sim0.8$ (aside from the primary peak at $C\sim0.5)$. This is consistent with a bimodal distribution of mass ratio for this event. For both events, the recovered mass ratio is consistent with the corresponding BBH analyses, despite the fact that~\bbh~ (and therefore,~\deco~) does not include higher multipoles. Taken together with the broader event sample, these cases provide no evidence for a clear trend in which mass-ratio effects alone drive significant displacements of the compactness posterior away from $C\simeq0.5$.
%
%
\subsection{Events from O4}

For this first study, where we establish our methods, we have chosen to focus on GWTC-3, and will consider the complete set of published observations in future work. However, we do analyze a couple of of O4 events that are of particular interest because of high SNR or unusual source properties.

GW250114 is the loudest GW event observed to date, with a network SNR of $\sim80$. The loudness of the event makes it a particularly valuable target for searches for departures from the BBH expectation; indeed, it has already been used for precision tests of GR and BH spectroscopy~\cite{LIGOScientific:2025wao,Akyuz:2025seg,Grimaldi:2026prn}, all of which find consistency with GR predictions. 
For GW250114, we find that the low-compactness mode seen in some GWTC-3 events also appears when analysing the Livingston data from the standard low-frequency cutoff. Its recurrence in such a high-SNR event reinforces that this failure mode is not simply a consequence of low signal strength. As in the GWTC-3 cases, raising the Livingston low-frequency cutoff to 50 Hz removes the low-compactness support and recovers BBH-like behaviour, with $C_{\rm eff}\sim0.52^{+0.01}_{-0.01}$ at $90\%$ CI. 

\begin{figure}[t!]
	\includegraphics[width=0.46\textwidth]{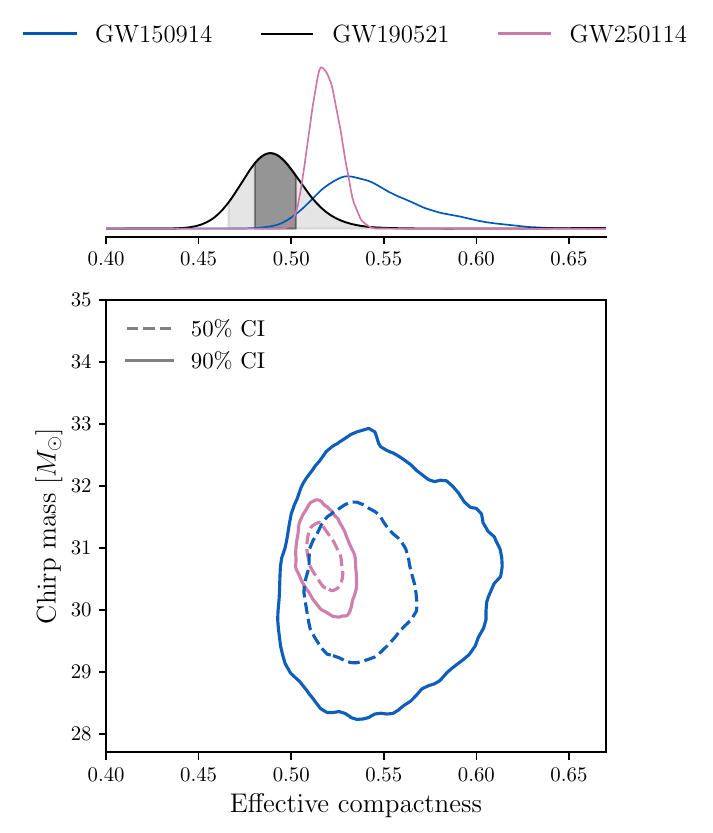}
	\caption{ The lower panel shows the two-dimensional posterior distributions for GW150914, in blue, and GW250114, in pink. Solid contours denote the 90\% CI, while dashed contours denote the 50\% CI. Although the two events are otherwise very similar, the higher SNR of GW250114 yields substantially tighter posteriors and shifts the posterior peak toward $C_{\rm eff}\sim 0.5$. The upper panel shows the one-dimensional posterior on effective compactness. For comparison, we include GW190521, which provided the best-constrained effective-compactness posterior in GWTC-3, with an SNR$\sim15$. The dark shaded region marks the 50\% CI and the lighter shaded region marks the 90\% CI. The full one-dimensional posteriors for GW250114 and GW150914 are also shown for comparison. 
	}
\label{fig:gw250114}
\end{figure}

Figure~\ref{fig:gw250114} places this final BBH-like posterior in context by comparing GW250114 with GW150914, a similar but much lower-SNR event. The comparison shows that the loudness of GW250114 substantially improves the compactness constraint (just as other parameters eg. chirp mass), not only by narrowing the posterior but also by shifting its peak closer to the expected BBH value, $C_{\rm eff}\simeq0.5$. The small residual offset from $C_{\rm eff}=0.5$ suggests that some further calibration of the phenomenological compactness mapping may be required to recover the BBH limit without bias. However, given the precision with which the merger morphology is measured, the complete absence of support at $C_{\rm eff}<0.5$ places an extremely strong constraint on the BH nature of GW250114, and also illustrates our ability to constrain compactness with even a simple phenomenological treatment. 

%
\subsection{GW231123 - a special case}
\label{sec:gw231123}


\subsubsection{Event properties and analysis challenges}
\label{sec:event_prop}
We consider here the inference of source properties of GW231123 with~\deco. This event is the most massive BBH system observed to date by the LVK and was detected by the LIGO Livingston and Hanford observatories. Because of the binary’s short in-band duration, the detected signal is dominated by the merger-ringdown phase, with only a very brief inspiral contribution. The data from both detectors were also affected by non-Gaussian noise transients at times close to the merger signal~\cite{GW231123}. Data-quality studies found that the transient in the Livingston data had no measurable impact on parameter estimation. However, glitches occurring in the temporal vicinity of the event in the Hanford detector may complicate the accurate inference of the source properties.

Aside from the large total mass, this event was also reported with a large component of misaligned spin, making it particularly challenging for state-of-the-art waveform approximants to converge on the source parameters. In fact it was demonstrated~\cite{GW231123,Bini:2026kwz} that similar systematics across waveform approximants could be reproduced even in the absence of noise, of course different noise realizations had different degree of impact on model systemics.

\subsubsection{Effective compactness inference}
\label{sec:inference}

\begin{figure}[h]
	\includegraphics[width=0.46\textwidth]{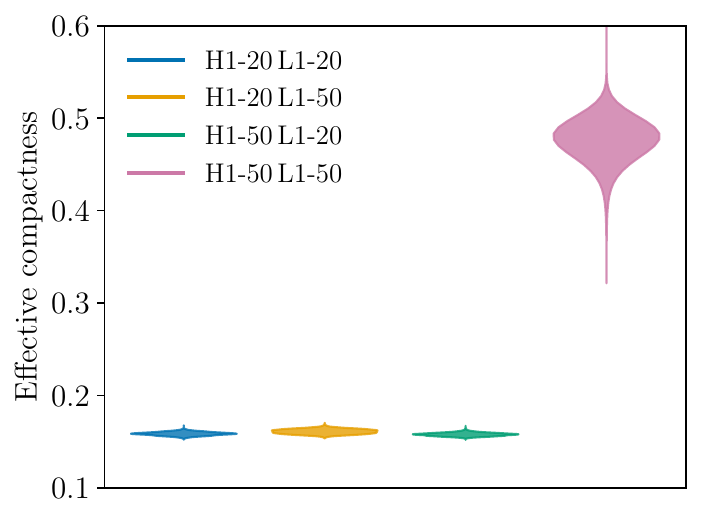}
	\caption{ Effective compactness inference for GW231123 using data above 20~Hz in both Livingston and Hanford is shown in blue. Raising the low-frequency cutoff to 50~Hz in either detector, while keeping the other at 20~Hz, leaves the compactness posterior largely unchanged, as shown in orange and green. A compactness posterior consistent with BBH, shown in pink, is recovered only when the low-frequency cutoff is set to 50~Hz in both detectors. }
\label{fig:gw231123}
\end{figure}

The application of~\deco~to the real GW231123 data produced an unexpected sequence of results. In the first three analysis configurations--using data above 20 Hz in both detectors, or raising the low-frequency cutoff to 50 Hz in only one detector at a time--the effective-compactness posterior showed pathologies similar to noted earlier in the low-compactness failure modes identified in the GWTC-3 sample. This motivated a final robustness test in which both Hanford and Livingston data were analysed only above 50 Hz. As shown in  Fig~\ref{fig:gw231123}, only for the final analysis do we recover a compactness consistent with expectations for BBHs and comparable to the compactness distribution inferred from GWTC-3. This behaviour differs from the low-compactness cases identified in GWTC-3, where raising the low-frequency cutoff in Livingston alone was sufficient to restore BBH-like compactness. For GW231123, the persistence of the low-$C$ mode under single-detector cutoff changes indicates that the effect is not isolated to one detector in the same way. 

\begin{figure*}[t]
	\includegraphics[width=0.94\textwidth]{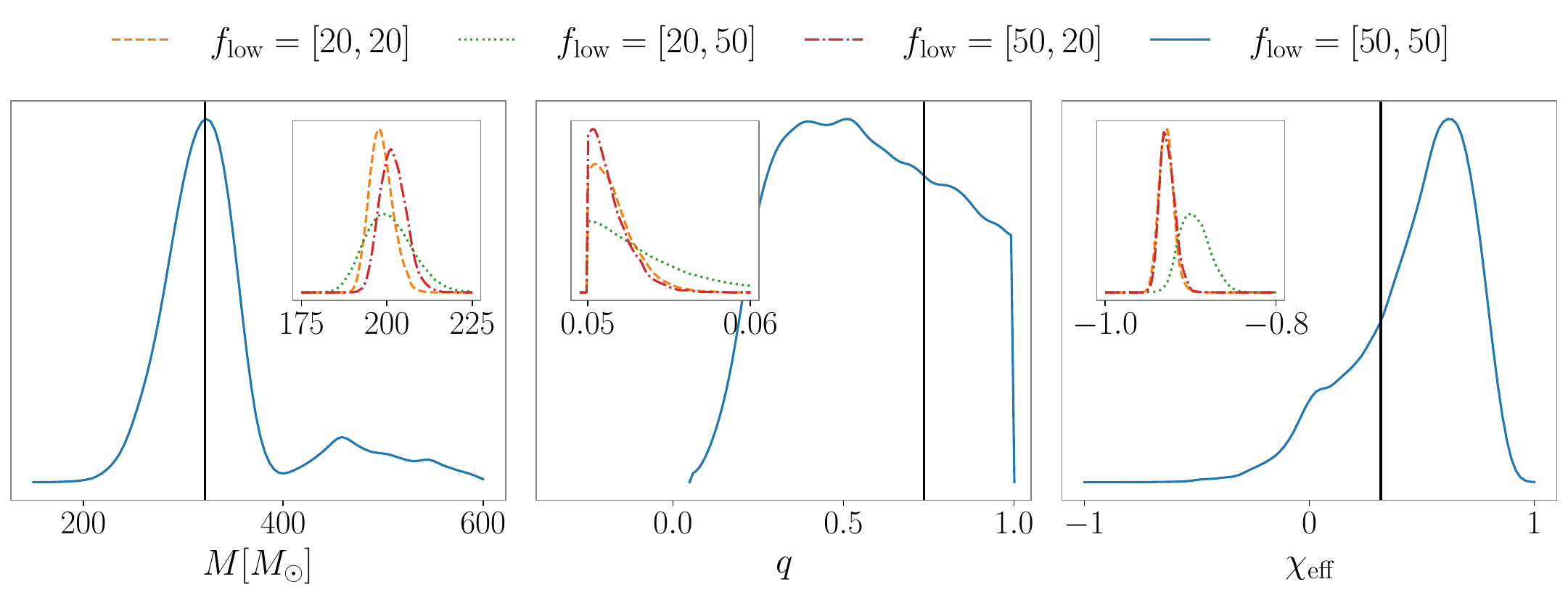}
	\caption{Posterior probability distributions for the total mass (\textit{Left}), mass ratio (\textit{Middle}), and effective spin (\textit{Right}), obtained when analyzing GW231123 with different low-frequency cut-offs in the Hanford and Livingston detectors. The solid vertical lines indicate the median values of the combined posterior distributions obtained in the LVK analysis, as reported in~\cite{GW231123}. The insets in all panels show a zoomed in region of the parameter space.}
\label{fig:gw231123_posteriors}
\end{figure*}

Figure~\ref{fig:gw231123_posteriors} compares the posterior distributions for the total mass, mass ratio, and effective spin inferred with~\deco~against the median of combined posteriors across several analyses reported in~\cite{GW231123}. Among these, the analysis using a 50~Hz low-frequency cutoff gives estimates that are most consistent with those reported in~\cite{GW231123}.

\subsubsection{~\deco~at high masses}
\label{sec:systematics}

The data-quality complications discussed in Sec.~\ref{sec:event_prop}, together with the modified frequency treatment required in Sec.~\ref{sec:inference} to obtain non-pathological inference with~\deco, motivate a closer examination of the model in this regime. We therefore verify that~\deco~can robustly infer source parameters for very high-mass, short-duration systems, including cases with significant spin precession. The short in-band duration of very massive binaries raises an important concern for~\deco~ inference in assigning low effective compactness simply because the model can make the waveform shorter by shifting the IMR transitions to lower frequencies. In this case, an apparent low-compactness posterior would not reflect non-BBH merger morphology, but rather a failure mode of the parametrization in the high-mass regime. We first test this possibility with a high-mass nonspinning BBH injection generated using~\bbh~and recovered with~\deco~in zero noise, which isolates the effect of total mass and signal duration from additional complications such as spin precession. In these control cases, ~\deco~recovers the injected source parameters and returns compactness posteriors consistent with the BBH expectation, indicating that high total mass alone does not generically produce a spurious low-compactness measurement.

\begin{figure}[h]
	\includegraphics[width=0.46\textwidth]{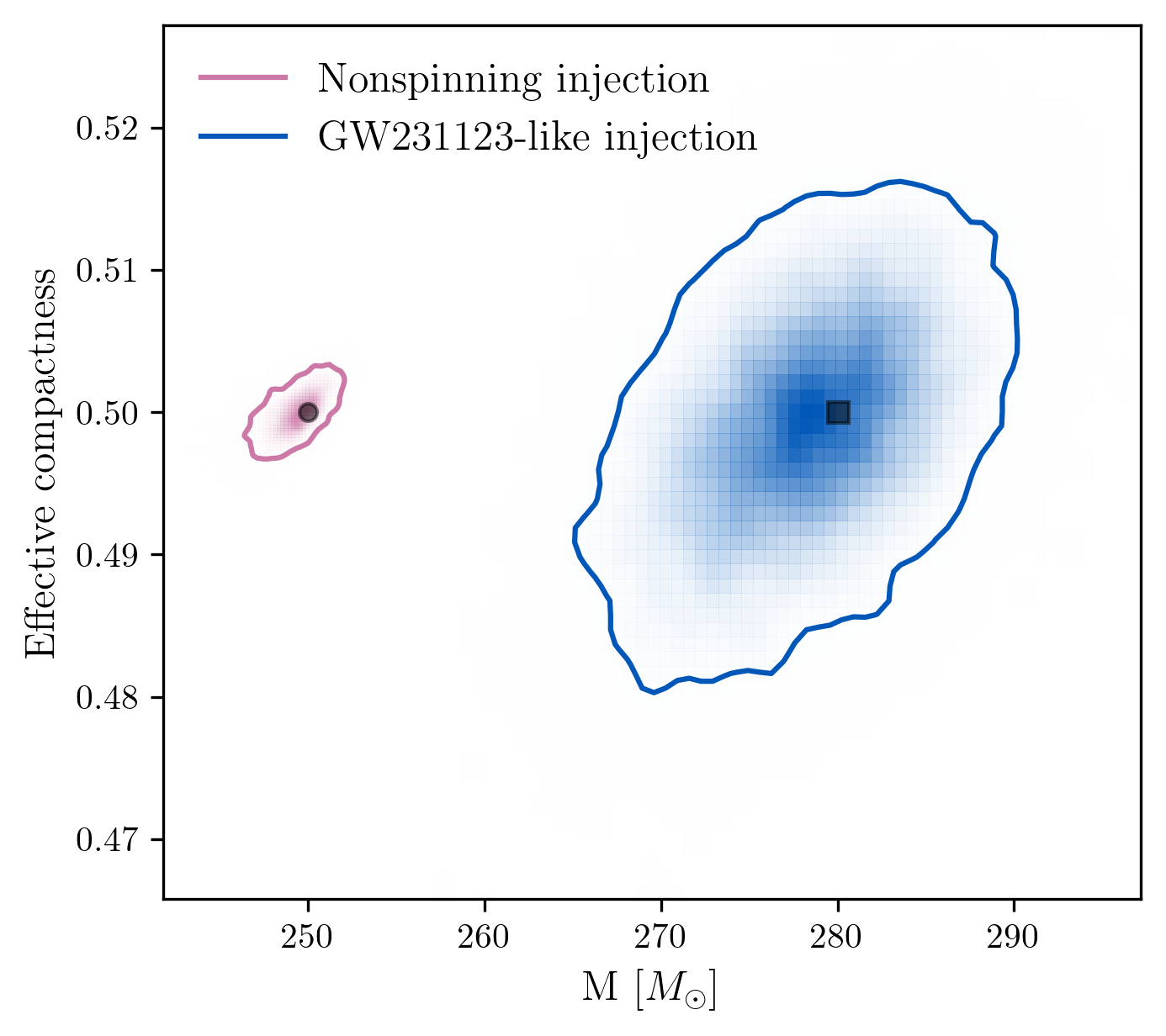}
	\caption{Two-dimensional posterior distributions in total mass and effective compactness for high-mass BBH injections recovered with~\deco. The pink contour shows the $90\%$ credible region for a high-mass nonspinning control injection with $M=250~\msun$, while the blue contour shows the corresponding region for a high mass precessing injection with $M=280~\msun$. The injected values, with effective compactness fixed to the BBH value $C_{\rm eff}=0.5$, are indicated by a solid circle (square) for the nonspinning (precessing) injections. While $\chi_{\rm{eff}}=0$ is recovered well in the nonspinning case, the injected $\chi_{\rm{eff}}=0.34$ is completely excluded at  $90\%$ CI in the precessing case, without noticeable biases in mass ratio for either case.}
\label{fig:highmass}
\end{figure}

We then repeat the test with a more challenging injection designed to resemble GW231123. Specifically, we inject a~\xphm~waveform in zero noise at SNR $\sim 26$, using the upper limits on the inferred component masses and the median values of $(\chi_{\rm eff}, \chi_p, D_L, \theta_{jn})$ where $D_L$ and $\theta_{jn}$ are the luminosity distance and inclination to source, respectively. Despite the short duration and the large misaligned-spin component,~\deco~again recovers the source parameters and yields an effective compactness consistent with the BBH value. Figure~\ref{fig:highmass} shows the recovery for this injection, along with the nonspinning high-mass BBH injection and, neither of them show noticeable biases in the recovered $M$, $q$ or $C_{\rm eff}$. $\chi_{\rm eff}=0.34$ is excluded at the $90\%$ CI, consistent with biases in $\chi_{\rm eff}$ seen for precessing systems with~\deco~(cf. fig.~\ref{fig:systematics}). These injection studies therefore indicate that neither high mass nor misaligned spin drives~\deco~toward a spurious low-compactness posterior, either individually or in combination, at least in the absence of detector noise. 

Additionally, we analyzed another short duration event, GW190521, that has been considered for several different exotic scenarios and found~\deco~ to favour a BBH classification. In fact, GW190521 yields the tightest constraint on compactness parameter across GWTC-3 with a clear preference for BBH, as depicted in the top panel of figure~\ref{fig:gw250114}. Notably, despite the lower SNR of this event ($\sim15$) compared to GW231123 ($\sim21$), we did not need to employ any non-standard settings in bayesian inference to recover a compactness consistent with BBH. This suggests that high total mass and large component of misaligned spin do not inherently preclude a BBH characterization using effective-compactness inference with~\deco~on real detector data. 

%

\subsubsection{Interpretation}

Within our analysis, GW231123 stands out as a special case in that a BBH-like characterization with~\deco~is obtained only after raising the low-frequency cutoff consistently in both detectors. Ordinarily, the persistence of a low-compactness mode under single-detector cutoff tests would make the event more suggestive of genuine non-BBH structure. However, GW231123 requires a more nuanced interpretation. Although the BBH-like result is recovered only with a modified low-frequency treatment, the analysis configurations that favour non-BBH compactness exhibit clear pathological behaviour. We therefore regard the BBH-like characterization as the most reliable among the runs considered here,

The BBH interpretation of GW231123 should nevertheless be caveated. As discussed above,~\deco~is designed to probe departures from BBH morphology primarily through the late-inspiral and merger transition. The interpretation of the compactness posterior therefore depends on whether this part of the waveform carries appreciable information in the data. Since the observed signal contains only about three cycles, the analysis may be driven either by late-inspiral/merger morphology or predominantly by ringdown. If sufficient late-inspiral and merger information remains in the data, then the recovery of BBH-like compactness when both detectors are analysed above 50 Hz provides meaningful support for a BBH-like merger morphology with~\deco.

On the other hand, if the observation is effectively dominated by ringdown alone, then a BBH-like characterization with~\deco~is less informative. The present version of~\deco~does not model an exotic post-merger signal; after the compactness-dependent inspiral--merger modification, it continues with a BBH-like ringdown. A model whose ringdown is always BH-like can therefore fit a BH remnant efficiently even if the components were not BHs. Nevertheless, this does not make the compactness result completely arbitrary. For a genuinely low-compactness binary that promptly forms a BH remnant, the lower compactness would shift the merger/contact scale to lower frequencies; matching the observed frequency content would generally require different source parameters, in particular a substantially larger total mass of the binary, which would then change the expected remnant ringdown frequency. The fact that the BBH-like~\deco~analysis recovers source properties broadly consistent with BBH analyses therefore gives some support to the standard interpretation, although it should not be read as a definitive exclusion of exotic scenarios. 

Additionally, the absence of flagged coincident data-quality issues, together with the fact that the anomaly depends on a common low-frequency cutoff across detectors, makes a purely uncorrelated detector-noise explanation less compelling than in the GWTC-3 cases. Our analysis with GW231123 therefore motivates targeted follow-up studies of shared low-frequency systematics, glitch mitigation, and non-CBC transient models before any non-standard astrophysical interpretation can be drawn.

\section{Waveform subtraction and residual diagnostics}
\label{sec:residuals}


The compactness posterior identifies events for which the data prefer merger morphologies away from the BBH expectation within the~\deco~parametrization. A complementary question is whether the waveform associated with the inferred compactness leaves residual structure in the data that is meaningfully different from that left by a standard BBH waveform. Residual-based diagnostics have also been proposed as a way to search for unmodeled physics in GW data: for example, the SCoRe framework looks for structured correlations in residual strain between detector pairs in order to distinguish noise artefacts from possible unmodeled signal components~\cite{Dideron:2022dcm}. In this section we therefore compare time--frequency residuals obtained after subtracting the maximum-likelihood~\deco~and BBH waveforms from the detector strain. This diagnostic is not intended as a detection statistic, an evidence calculation, or a replacement for Bayesian model selection, but rather as a complementary robustness check. A Bayes factor between BBH and~\deco~would be sensitive to the prior volume assigned to the phenomenological compactness parameter, including regions that do not yield a significant likelihood improvement over the BBH model and therefore contribute to the Occam penalty~\cite{Thrane_2019}. The residual comparison avoids this particular dependence on the compactness prior by conditioning on the maximum-likelihood BBH and~\deco~waveforms and comparing the residual time--frequency power left by each subtraction.

For a given analysis, we construct residual data by subtracting a detector-projected maximum-likelihood waveform from the publicly released calibrated strain. Specifically, for each detector, we take the maximum-likelihood waveform from the Bayesian analysis, project it onto the detector response, and construct the corresponding signal time series. We then place this signal on the same time grid as the publicly available strain time series and subtract it sample by sample from the detector strain~\cite{LIGOScientific:2021sio}. We repeat this procedure for both the baseline BBH waveform and~\deco~, thereby obtaining two residual time series for direct comparison. 

\subsection{Residual contrast as a detector-level diagnostic}

To assess the remaining time-frequency structure, we examine the Q-transform spectrogram of each residual~\cite{Chatterji}. The data are first whitened using the detector noise power spectral density estimated from data segments excluding the chirp signal. We then use the \texttt{q\_transform} method implemented in \texttt{gwpy} for \texttt{TimeSeries} objects~\cite{MACLEOD2021100657} to project it onto time-frequency tiles with varying duration and bandwidth. This spectrogram is computed over a fixed window of $\pm 0.4\,\mathrm{s}$ around the merger time and over the frequency band $20\text{--}500\,\mathrm{Hz}$. The normalized energy shown in the spectrogram measures the excess power in each tile relative to the local noise expectation.

Localized transient excess power is captured by a relatively small subset of time--frequency tiles with large normalized energy, whereas stationary noise is distributed more broadly across the tiling. Therefore, summing the normalized energies of tiles above a threshold provides a simple measure of the localized residual structure left after subtraction. Motivated by this, we define, for each detector separately, a diagnostic---\emph{residual contrast}---that compares the residuals obtained from the two waveform models, as
\begin{equation}
\Delta_{\mathrm{res}}(\tau) =\Biggl|\frac{\sum\limits_{i\in \tau} E^{\rm{BBH}}_{i}-\sum\limits_{i \in \tau}E^{~\deco~}_{i}}
     {\sum\limits_{i\in\tau} E^{\mathrm{strain}}_{i}} \Biggr|,
\end{equation}
where $E^{BBH/~\deco~}_{i}$ is the normalized energy in $i^{\rm{th}}$ tile of the residual spectrogram, $E^{\mathrm{strain}}_{i}$ is the normalized energy in the $i^{\rm{th}}$ tile of the original strain data, and the summation is carried out over tiles with normalized energy above the chosen threshold,~$\tau$. We use $\Delta_{\mathrm{res}}$ only as an event-level diagnostic, not as a detection statistic or cross-event ranking. It measures whether the BBH waveform and~\deco~maximum-likelihood subtractions leave comparable localized excess power in a given detector. Because the normalization is detector-specific, values should not be compared directly across events. We report the residual contrast for a subset of events in Table~\ref{tab:residual_contrast} and discuss three representative cases below to illustrate its interpretation.

\begin{table}[h]
    \centering
    \caption{Detector-level residual contrasts for selected events shown for a choice of threshold on normalized energies, only for the Livingston detector (LLO)}
    \label{tab:residual_contrast}
    \setlength{\tabcolsep}{4pt}
    \begin{tabular}{lccc}
        \hline
        Event & Detector & Threshold  & Residual contrast \\
        \hline
         GW150914 & LLO & 10 & 0.028 \\
         GW170608 & LLO & 10 & 0.013\\
         GW170729 & LLO & 10 & 0.012\\
         GW190408 & LLO & 10 & 0.016\\
         GW190412 & LLO & 10 & 0.013\\
	 GW190915 & LLO & 10 & 0.024\\
         GW200129 & LLO & 10 & 0.017\\
         GW19112 & LLO & 10 & 0.037\\
         GW200219 & LLO & 10 & 0.077\\
        \hline
    \end{tabular}
\end{table}

\subsubsection{GW200129}
\begin{figure*}[t]
	\includegraphics[width=0.84\textwidth]{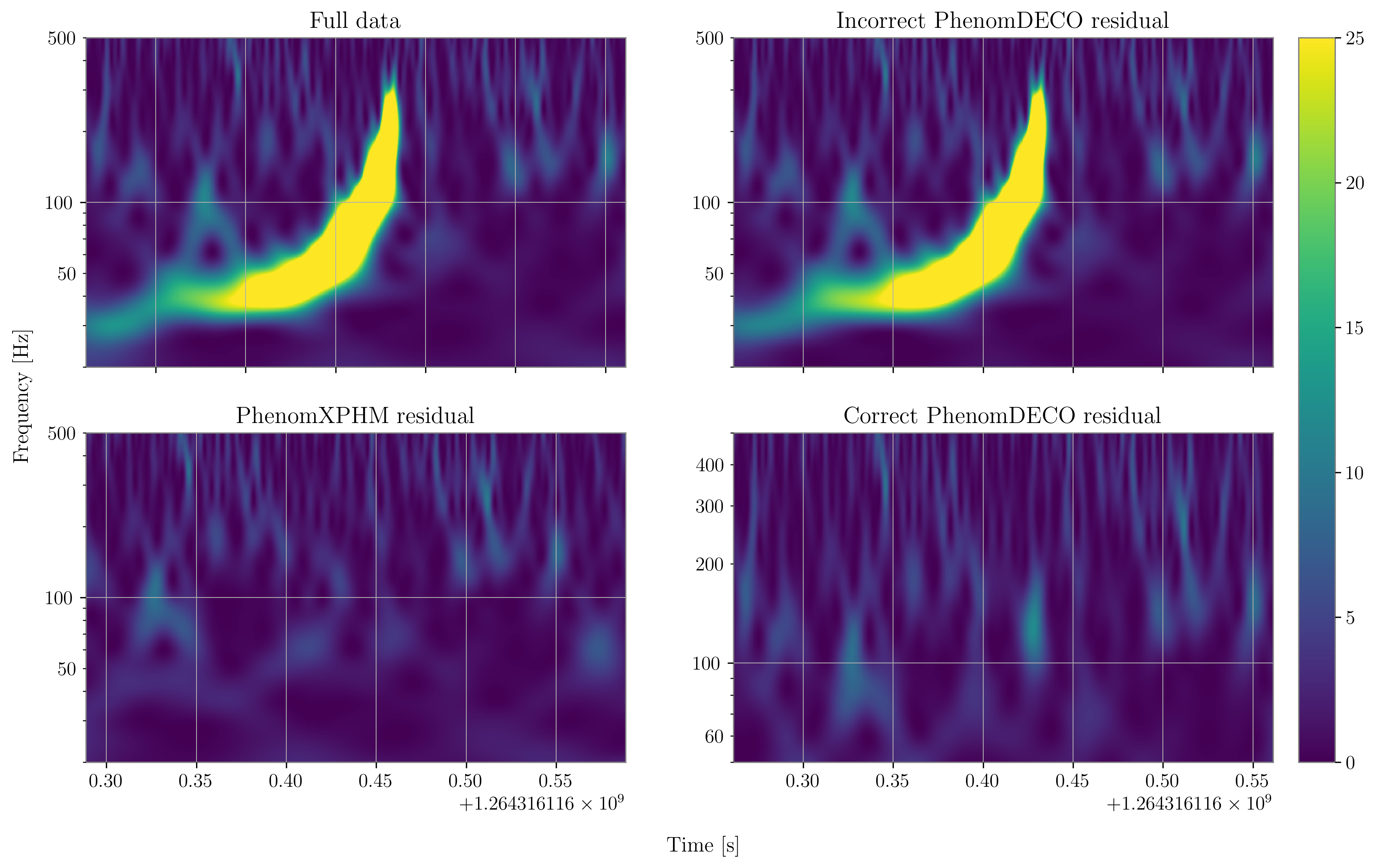}
	\caption{Top left: spectrogram of the publicly available deglitched strain data for GW200129. Top right: residual spectrogram obtained after subtracting the maximum-likelihood~\deco~waveform from the analysis that favors $C\sim0.15$. Bottom left: residual spectrogram obtained after subtracting the maximum-likelihood BBH waveform. Bottom right: residual spectrogram obtained after subtracting the maximum-likelihood~\deco~waveform from the analysis using a 50 Hz lower frequency cutoff for Livingston. }
\label{fig:residualsGW200129}
\end{figure*}

First, we consider GW200129 which is one of the events where we found a spurious low-compactness peak at $\sim0.15$. Using a low frequency cutoff of 50 Hz for Livingston data, while keeping 20 Hz for Hanford, shifts the recovered compactness posterior to a peak near 0.5. The associated spectrograms are consistent with this behaviour: the maximum-likelihood waveform from the analysis yielding the BBH-like compactness posterior leads to a clean subtraction and similar residual features as the standard BBH waveform (~\xphm~), as shown in Fig.~\ref{fig:residualsGW200129}
This provides an example where the compactness inference can be misleading if considered without residual diagnostics. In the nominal analysis, the compactness posterior develops a pronounced peak near $C \sim 0.15$, with no support in posterior probability distribution for $C=0.5$. In keeping with the erroneous geocentric trigger time, inspection of the residual spectrogram also shows that this low-compactness solution does not correspond to a clean subtraction of the chirp signal. When the analysis is repeated after excluding data below 50\,Hz in the Livingston detector, the compactness posterior shifts back to a BBH-like value near $C \sim 0.5$, and the corresponding residual spectrogram is much cleaner. This behaviour corroborates that the original low-compactness peak was spurious and driven by low-frequency artefacts rather than genuine non-BBH signal morphology.

\subsubsection{GW191127}
We then consider GW191127 and find the likelihood function to peak for a compactness $C \sim 0.35$ ($\rm{log} \mathcal{L}^{C=0.5}=33.69$ and $\rm{log} \mathcal{L}^{C=0.35}=48.06$ ) in Bayesian analysis, with a residual contrast of 0.037 between~\deco~and~\xphm~. This is comparable to what we find for events in which the likelihood peaks near $C\sim0.5$ (see Table~\ref{tab:residual_contrast}). Taken in isolation, this might suggest that a moderately reduced compactness provides an equally good description of the data. However, for this event the maximum-likelihood (maxL) total mass and mass ratio are shifted relative to the BBH analysis, as expected from the degeneracies discussed in Ref.~\cite{Ghosh:2025wex}. In other words, the improved agreement is achieved only by compensating the modified merger morphology with biased intrinsic parameters as shown in Table~\ref{tab:gw191127}.
\begin{table}[h]
    \centering
    \caption{Total mass and mass ratio corresponding to the maximum likelihood obtained in Bayesian analysis with~\deco~and~\xphm~ for GW191127.}
    \label{tab:gw191127}
    \setlength{\tabcolsep}{6pt}
    \begin{tabular}{cccc}
        \hline
        Waveform & max(log $\mathcal{L}$) & $M^{\rm{maxL}}$  & $q^{\rm{maxL}}$ \\
        \hline
         ~\deco~ & 48.06 & 52.5 \msun & 0.24\\
         \xphm~ & 45.38 &72.1 \msun & 0.77\\
        \hline
    \end{tabular}
\end{table}
This highlights that a clean subtraction is not sufficient to support an exotic interpretation and supports our conservative requirement that $C=0.5$ lie outside the 90\% CI of the compactness posterior before treating an event as a candidate departure from the BBH hypothesis.

\subsubsection{GW200219}
\begin{figure*}[t]
    \centering
    \begin{subfigure}{0.45\linewidth}
        \includegraphics[width=\linewidth]{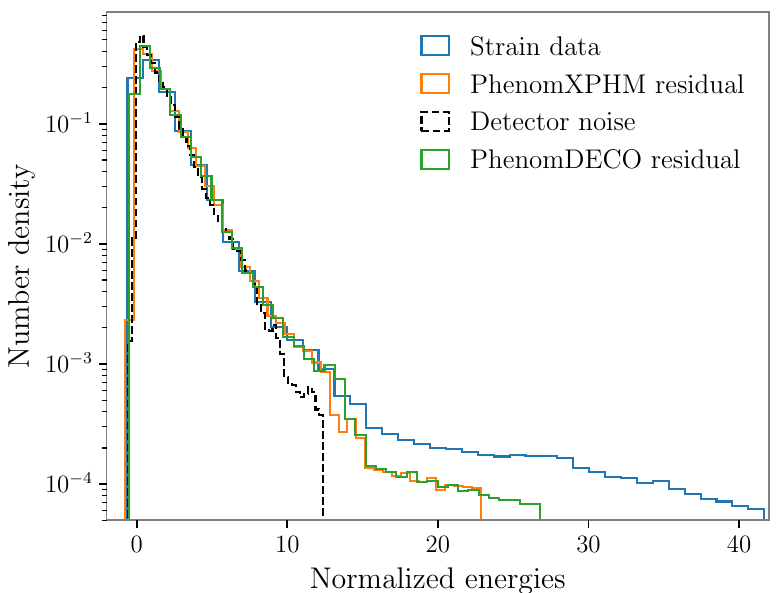}
    \end{subfigure}
    \hfill
    \begin{subfigure}{0.45\linewidth}
        \includegraphics[width=\linewidth]{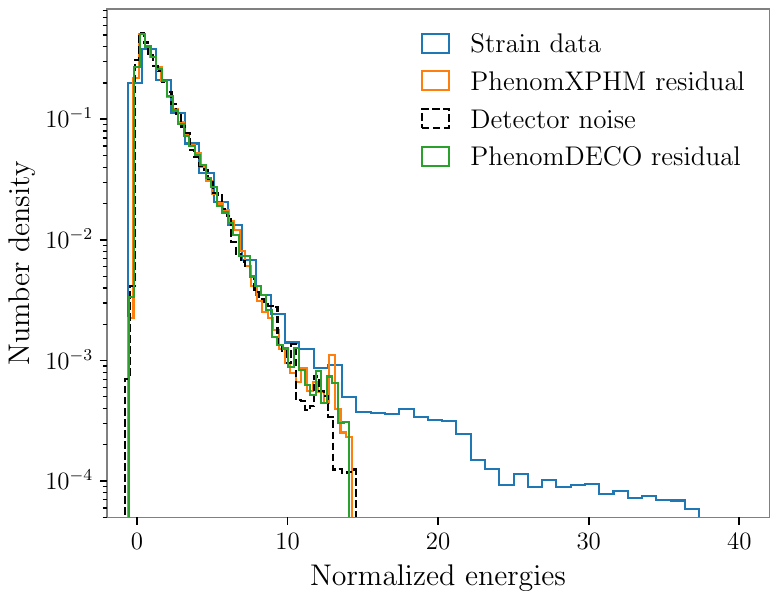}
    \end{subfigure}
    \caption{Normalized energy distributions for two events. Left panel shows GW200219, which was identified in LVK tests of GR as having relatively large residual power after subtraction of the best-fit GR waveform. Right panel shows GW190408 as a comparison case with a cleaner BBH subtraction. In both panels, the~\xphm~ and~\deco~residual distributions are compared with the original strain data and with the detector-noise distribution estimated from data excluding the signal. The residuals for GW190408 lie closer to the detector-noise distribution than those for GW200219, illustrating the difference between a poorer and a cleaner BBH subtraction.}
        \label{fig:residual_histogram}
\end{figure*}

Finally, we consider GW200219, which was reported with the largest residual SNR and, more importantly, the poorest fit by the GR template in LVK GWTC-3 tests of GR analysis($\rm{FF}_{\rm{90}}=0.74$ see table III in~\cite{LIGOScientific:2021sio} for details). Our analysis yields a compactness posterior consistent with the BBH expectation, with a maxL value of $C^{\rm maxL}\sim 0.482$, and a residual contrast of 0.077, comparable to events in our sample where likelihood peaks at $C\sim0.5$. Left panel on Fig.~\ref{fig:residual_histogram} shows the distribution of normalized energies in the spectrograms for the Livingston strain, detector noise and residuals after subtraction with~\xphm~ and~\deco~\cite{Liang:2025zws}. For comparison, right panel on Fig.~\ref{fig:residual_histogram} shows the distribution of normalized energies for GW190408\_181802 in which the BBH template provides a much cleaner subtraction ($\rm{FF}_{\rm{90}}=0.88$). The residual distributions in this case lie noticeably closer to the detector-noise distribution, which is obtained from the spectrogram of the detector data excluding the signal.

Notably, the residual contrast does not increase drastically for nearby lower-likelihood samples, for example at $C=0.4$, $\Delta_{\rm{res}}=0.072$. By contrast, if the compactness is artificially reduced to $C\sim 0.3$ (which has no support in the posterior probability distribution of the compactness parameter) while the remaining source parameters are held fixed at their maxL values, the subtraction deteriorates visibly with a residual contrast of $\Delta_{\rm{res}}=0.72$. 

It is worth emphasizing that the residual contrast is primarily a measure of whether the two waveform models leave comparable residual structure, in which case it is expected to be close to zero. It is not, however, a quantitative statistic for comparing subtraction quality across different events, since its normalization is set by the total strain spectrogram energy in a given detector and therefore varies from event to event. Its usefulness instead lies in interpreting, on an event-by-event basis, whether anomalous residual structure is better accounted for by the specific waveform deformation encoded by~\deco~. In this sense, it is complementary to the LVK residual analysis, which identifies events for which the best-fit BBH waveform leaves anomalous residual power, whereas the~\deco~residual analysis tests whether such an anomaly is consistent with a physically targeted deformation of the merger morphology. At least in the case of GW200219, we find that, despite the moderately poor fit of the GR template, this specific non-BBH deformation is not favoured.

We do not apply this diagnostic to GW231123 here, since its unusually complex time--frequency morphology would require a more tailored treatment than our simple thresholded residual-contrast statistic (cf. Fig.~1 of Ref.~\cite{GW231123}).

%
%

%
\section{Conclusion}

BBHs are the standard interpretation for the majority of compact binary mergers observed in GWs, but in practice this classification is driven largely by the inferred component masses inferred from data with waveform models that describe BBHs in GR. There are existing tests that probe the nature of compact objects by tracking deviations either in the inspiral, through tidal and spin-induced finite-size effects, or in the remnant, through ringdown consistency and post-merger searches. The merger morphology itself provides a complementary diagnostic, especially for high-mass or otherwise unusual systems in which much of the observed signal power lies near merger. This is the regime probed by~\deco~.

In this work, we have presented the first systematic measurement of the effective compactness parameter inferred by~\deco~across a large sample of gravitational-wave events. Using the GWTC-3 BBH population as a benchmark, we find that most events are consistent with the BBH expectation, with compactness posteriors peaked near or including $C=0.5$. We also identify two recurring classes of non-standard posterior behaviour: events with support at $C>0.5$, which we interpret as BBH-like within the present parametrization rather than as evidence for a physical compactness above the black-hole value, and events with sharp low-compactness modes near $C\simeq0.15$. The latter are accompanied by additional signs of unreliable inference, including poor recovery of the chirp mass, anomalously low single-detector SNRs, and weak localization of the merger time. They are also not robust under changes to the low-frequency analysis cutoff: in all such GWTC-3 cases, raising the Livingston cutoff removes the low-compactness mode and restores BBH-like compactness. This population-level behaviour is summarized in Fig.~\ref{fig:full}, where the compactness posteriors are shown on a common set of bins. After the Livingston reanalysis, the events that initially showed sharp support near $C\simeq0.15$ no longer stand out as low-compactness outliers. 

GW231123 also displays a low-compactness mode when analyzed from 20 Hz, but its behavior is qualitatively different from the low-C cases identified in the GWTC-3 benchmark. In the GWTC-3 events, the low-C mode was removed by raising the low-frequency cutoff in Livingston alone, indicating a detector-specific low-frequency artifact. For GW231123, by contrast, the low-C mode persists when the cutoff is raised in only one detector, and BBH-like compactness is recovered only when both Hanford and Livingston are analyzed above 50 Hz. This distinction makes GW231123 an outlier with respect to the empirical benchmark established here. The result does not by itself support an exotic interpretation: the event is high-mass, short-duration, and affected by known data-analysis challenges, and apparent deviations in merger morphology can arise from noise structure, waveform systematics, or missing physics in the model. However, the absence of flagged coincident data-quality issues, together with the fact that the anomaly depends on a common low-frequency cutoff across detectors, makes a purely uncorrelated detector-noise explanation less compelling than in the GWTC-3 cases. GW231123 therefore motivates targeted follow-up studies of shared low-frequency systematics, glitch mitigation, and non-CBC transient models before any non-standard astrophysical interpretation can be drawn.

Beyond the compactness-based criteria of Sec.~\ref{sec:exotic_def}, it is important to examine whether the preferred~\deco~solution leaves residual structure in the data that is meaningfully different from a BBH subtraction. For example, for GW191127 the likelihood peaks at $C\simeq0.35$, and the residual contrast between the maximum-likelihood~\deco~and BBH subtractions is comparable to that found in events where the likelihood instead peaks near $C\simeq0.5$. Taken alone, this could suggest that a moderately lower compactness provides an equally viable description of the data. However, the maximum-likelihood total mass and mass ratio inferred with ~\deco~are not consistent with those obtained in the BBH analysis, indicating that the lower-compactness solution is accompanied by compensating shifts in the intrinsic parameters rather than a cleanly isolated change in merger morphology. If $C=0.5$ were confidently excluded this event would have been interesting for follow-up but in the present analysis the broad compactness posterior prevents such an interpretation. This also highlights the importance of reducing these degeneracies, for example through improved~\deco~modelling that narrows the compactness posterior, will be essential for turning such cases into robust candidate tests of non-BBH merger morphology.

GW200219 provides a complementary example of the role of residual diagnostics. This event was identified in the LVK tests of GR as having unusually large residual power after subtraction of the best-fit GR waveform, indicating that the GR template was a poor description of the data. In those tests, the residual analysis searches for coherent excess power remaining in the stretch of data analysed after the CBC signal has been removed, thereby asking whether the data contain structure not captured by the best-fit BBH model. The residual diagnostic used here asks a more targeted question: whether the residual power is better accounted for by the specific parametric deformation of the late-inspiral to merger morphology introduced by~\deco~. For GW200219, we find a compactness posterior consistent with the BBH expectation and the residual histograms to be consistent with that of detector noise. Thus, although GW200219 shows anomalous residual power in a generic residual test, that residual structure is not explained by the modification to merger morphology considered here. This highlights the complementarity of the two residual approaches: LVK-style residual tests can identify events with excess coherent power, while model-based residual analysis can test whether such excess power is consistent with a particular non-BBH modification of the merger morphology.

Across the GWTC-3 events considered here, we find no case that satisfies our robustness criteria for a non-BBH classification. Apparent low-compactness modes are removed once the detector-dependent low-frequency data treatment is varied, indicating that these features are more naturally explained by noise artefacts and inference pathologies than by genuine non-BBH merger morphology. GW231123 remains a special case: a BBH-like compactness posterior is recovered only after applying a higher low-frequency cutoff in both detectors, but controlled high-mass and precessing BBH injections suggest that neither high total mass nor spin precession alone drives~\deco~toward spurious low compactness. Overall, these results show that low-frequency data treatment is crucial before attributing apparent deviations from BBH expectations to exotic physics.

Taken together, these results establish effective compactness as a diagnostic of merger morphology and, through it, as a consistency check of the BBH interpretation. By applying~\deco~across a broad event sample, identifying recurring posterior morphologies, and testing the stability of apparent low-C support under detector-dependent low-frequency cutoffs, we have established a practical robustness framework for interpreting compactness inference on real GW data. Future work should focus on improving the~\deco~model in ways that reduce parameter biases and sharpen compactness measurements across events. This includes extending the waveform beyond its current amplitude-only aligned-spin implementation and incorporating relevant finite-size effects, such as tidal contributions, where appropriate. As the GW catalog continues to grow, the identification of compact binaries will become an increasingly empirical problem. A scalable method such as~\deco, which can be applied broadly across events as a first-pass BBH-consistency test, provides a practical way to identify systems that merit more targeted follow-up.

\section{Acknowledgements}

We thank N. V. Krishnendu for comments during the LIGO--Virgo--KAGRA internal review. SG thanks Erik Katsavounidis for valuable feedback on the presentation of residuals, and Salvatore Vitale, Gabriela Gonz\'alez and Emanuele Berti for insightful discussions. SG and FO acknowledges support from the Max Planck Society and CH thanks the University of Portsmouth for support through the Dennis Sciama Fellowship. MH was supported by Science and Technology Facilities Council (STFC) grant ST/V00154X/1. We are grateful for computational resources provided by the LIGO Laboratory, supported by the National Science Foundation Grants PHY-0757058 and PHY-0823459, and the SCIAMA high performance computing cluster supported by the Institute of Cosmology and Gravitation (ICG) and the University of Portsmouth.

This research has made use of data or software obtained from the Gravitational Wave Open Science Center (gwosc.org), a service of the LIGO Scientific Collaboration, the Virgo Collaboration, and KAGRA. This material is based upon work supported by NSF's LIGO Laboratory which is a major facility fully funded by the National Science Foundation, as well as the Science and Technology Facilities Council (STFC) of the United Kingdom, the Max-Planck-Society (MPS), and the State of Niedersachsen/Germany for support of the construction of Advanced LIGO and construction and operation of the GEO600 detector. Additional support for Advanced LIGO was provided by the Australian Research Council. Virgo is funded, through the European Gravitational Observatory (EGO), by the French Centre National de Recherche Scientifique (CNRS), the Italian Istituto Nazionale di Fisica Nucleare (INFN) and the Dutch Nikhef, with contributions by institutions from Belgium, Germany, Greece, Hungary, Ireland, Japan, Monaco, Poland, Portugal, Spain. KAGRA is supported by Ministry of Education, Culture, Sports, Science and Technology (MEXT), Japan Society for the Promotion of Science (JSPS) in Japan; National Research Foundation (NRF) and Ministry of Science and ICT (MSIT) in Korea; Academia Sinica (AS) and National Science and Technology Council (NSTC) in Taiwan. This material is based upon work supported by NSF's LIGO Laboratory which is a major facility fully funded by the National Science Foundation. For the purpose of open access, the author(s) has applied a Creative Commons Attribution (CC BY) licence to any Author Accepted Manuscript version arising.

\appendix
\section{Events table}
\label{app:appendix1}
\begin{table*}[t]
\centering
\caption{Median values of source-frame total mass $M$, the mass ratio $q \equiv m_2/m_1 \le 1$, and the network matched-filter SNR from Refs.~\cite{Abbott:2023gwtc21,Abbott:2021tsj} for the events analyzed in this work, with the index used in figure~\ref{fig:full}. }
\label{tab:event_catalog_summary}
\setlength{\tabcolsep}{7pt}
\begin{tabular}{c l c c c}
\hline
Index & Event & $M\,[\msun]$ & $q$ & $\rho_{\rm network}$ \\
\hline
0  & GW150914         & 64.5  & 0.87 & 26.0 \\
1  & GW151012         & 38.8  & 0.55 & 9.3  \\
2  & GW151226         & 21.7  & 0.53 & 12.7 \\
3  & GW170104         & 49.6  & 0.72 & 13.8 \\
4  & GW170608         & 18.5  & 0.74 & 15.3 \\
5  & GW170729         & 84.4  & 0.55 & 10.7 \\
6  & GW170809         & 58.5  & 0.71 & 12.8 \\
7  & GW170814         & 56.0  & 0.81 & 17.7 \\
8  & GW170818         & 62.5  & 0.79 & 12.0 \\
9  & GW170823         & 67.0  & 0.76 & 12.2 \\
10 & GW190408\_181802 & 34.3  & 0.64 & 14.6 \\
11 & GW190412         & 36.8  & 0.32 & 19.8 \\
12 & GW190413\_052954 & 58.0  & 0.72 & 9.0  \\
13 & GW190413\_134308 & 81.3  & 0.59 & 10.6 \\
14 & GW190421\_213856 & 73.6  & 0.76 & 10.7 \\
15 & GW190503\_185404 & 69.4  & 0.69 & 12.2 \\
16 & GW190512\_180714 & 35.8  & 0.54 & 12.7 \\
17 & GW190513\_205428 & 54.4  & 0.51 & 12.5 \\
18 & GW190514\_065416 & 69.3  & 0.69 & 8.0  \\
19 & GW190517\_055101 & 64.1  & 0.61 & 10.8 \\
20 & GW190519\_153544 & 105.6 & 0.63 & 15.9 \\
21 & GW190521         & 153.1 & 0.58 & 14.3 \\
22 & GW190521\_074359 & 76.3  & 0.77 & 25.9 \\
23 & GW190527\_092055 & 58.1  & 0.62 & 8.0  \\
24 & GW190602\_175927 & 115.6 & 0.62 & 13.2 \\
25 & GW190620\_030421 & 92.7  & 0.60 & 12.1 \\
26 & GW190630\_185205 & 59.4  & 0.68 & 16.4 \\
27 & GW190701\_203306 & 94.3  & 0.75 & 11.2 \\
28 & GW190706\_222641 & 112.6 & 0.67 & 13.4 \\
29 & GW190707\_093326 & 20.1  & 0.65 & 13.1 \\
30 & GW190708\_232457 & 31.4  & 0.59 & 13.4 \\
31 & GW190719\_215514 & 57.2  & 0.54 & 7.9  \\
32 & GW190720\_000836 & 21.8  & 0.53 & 10.9 \\
33 & GW190725\_174728 & 18.3  & 0.53 & 9.1  \\
34 & GW190727\_060333 & 68.8  & 0.78 & 11.7 \\
\hline
\end{tabular}
\hspace{0.5cm}
\begin{tabular}{c l c c c}
\hline
Index & Event & $M\,[\msun]$ & $q$ & $\rho_{\rm network}$ \\
\hline
35 & GW190728\_064510 & 20.7  & 0.64 & 13.1 \\
36 & GW190731\_140936 & 70.7  & 0.69 & 8.8  \\
37 & GW190803\_022701 & 65.0  & 0.73 & 9.3  \\
38 & GW190805\_211137 & 76.7  & 0.66 & 8.1  \\
39 & GW190828\_063405 & 57.2  & 0.81 & 16.5 \\
40 & GW190828\_065509 & 34.3  & 0.44 & 10.2 \\
41 & GW190910\_112807 & 78.0  & 0.78 & 14.5 \\
42 & GW190915\_235702 & 57.2  & 0.75 & 13.1 \\
43 & GW190924\_021846 & 13.9  & 0.58 & 12.0 \\
44 & GW190925\_232845 & 36.7  & 0.75 & 9.7  \\
45 & GW190929\_012149 & 93.3  & 0.40 & 9.7  \\
46 & GW190930\_133541 & 21.2  & 0.49 & 9.7  \\
47 & GW191103\_012549 & 20.0  & 0.67 & 8.9  \\
48 & GW191105\_143521 & 18.5  & 0.66 & 9.7  \\
49 & GW191109\_010717 & 112.0 & 0.72 & 17.3 \\
50 & GW191127\_050227 & 80.0  & 0.45 & 9.2  \\
51 & GW191129\_134029 & 17.5  & 0.63 & 13.1 \\
52 & GW191204\_171526 & 20.2  & 0.69 & 17.5 \\
53 & GW191215\_223052 & 43.3  & 0.73 & 11.2 \\
54 & GW191216\_213338 & 19.8  & 0.64 & 18.6 \\
55 & GW191222\_033537 & 79.0  & 0.77 & 12.5 \\
56 & GW191230\_180458 & 86.0  & 0.75 & 10.4 \\
57 & GW200112\_155838 & 63.9  & 0.79 & 19.8 \\
58 & GW200128\_022011 & 75.0  & 0.77 & 11.0 \\
59 & GW200129\_065458 & 63.4  & 0.84 & 26.8 \\
60 & GW200202\_154313 & 17.6  & 0.72 & 10.8 \\
61 & GW200208\_130117 & 65.4  & 0.72 & 10.8 \\
62 & GW200209\_085452 & 62.6  & 0.76 & 9.6  \\
63 & GW200219\_094415 & 65.0  & 0.74 & 10.7 \\
64 & GW200224\_222234 & 72.2  & 0.81 & 20.0 \\
65 & GW200225\_060421 & 33.5  & 0.73 & 12.5 \\
66 & GW200302\_015811 & 57.8  & 0.53 & 10.8 \\
67 & GW200311\_115853 & 61.9  & 0.81 & 17.8 \\
68 & GW200316\_215756 & 21.2 & 0.60 & 10.3 \\
\hline
\end{tabular}
\end{table*}
\newpage
\bibliography{refs.bib}
\end{document}